\documentclass[reprint,onecolumn,notitlepage,showpacs,preprintnumbers,nofootinbib,amsmath,amssymb,aps,prd,floatfix,superscriptaddress]{revtex4-1}
\pdfoutput=1
\usepackage{graphicx}
\usepackage{bm}
\usepackage{subfigure}
\usepackage{url}
\usepackage[hyperindex]{hyperref}
\usepackage{color}
\usepackage[ddmmyy,24hr]{datetime}
\usepackage{bigdelim}
\usepackage{booktabs}
\usepackage{dcolumn}
\usepackage{multirow}
\usepackage{subfigure}
\usepackage{cancel}
\usepackage{stackrel}
\usepackage{paralist}
\usepackage{xspace}
\usepackage{slashed}
\newcommand{\nua}[1]{\ensuremath{\rlap{\kern-2.5pt\ensuremath{\overset{\scriptscriptstyle(-)}{\phantom{\nu}}}}{\ensuremath{{\nu}_{#1}}}}}
\newcommand{\vet}[1]{\ensuremath{\hskip-1pt\vec{\hskip1pt#1}}}
\usepackage{enumitem}

\newcommand{\bluechange}[1]{{\color{black} #1}}

\newcommand{\QW}{$Q_{W}$}

\begin{document}

\title{Neutrino, Electroweak and Nuclear Physics from COHERENT Elastic Neutrino-Nucleus Scattering with a New Quenching Factor}

\author{M. Cadeddu}
\email{matteo.cadeddu@ca.infn.it}
\affiliation{Dipartimento di Fisica, Universit\`{a} degli Studi di Cagliari,
and
INFN, Sezione di Cagliari,
Complesso Universitario di Monserrato - S.P. per Sestu Km 0.700,
09042 Monserrato (Cagliari), Italy}

\author{F. Dordei}
\email{francesca.dordei@cern.ch}
\affiliation{Istituto Nazionale di Fisica Nucleare (INFN), Sezione di Cagliari,
Complesso Universitario di Monserrato - S.P. per Sestu Km 0.700,
09042 Monserrato (Cagliari), Italy}

\author{C. Giunti}
\email{carlo.giunti@to.infn.it}
\affiliation{Istituto Nazionale di Fisica Nucleare (INFN), Sezione di Torino, Via P. Giuria 1, I--10125 Torino, Italy}

\author{Y.F. Li}
\email{liyufeng@ihep.ac.cn}
\affiliation{Institute of High Energy Physics,
Chinese Academy of Sciences, Beijing 100049, China}
\affiliation{School of Physical Sciences, University of Chinese Academy of Sciences, Beijing 100049, China}

\author{Y.Y. Zhang}
\email{zhangyiyu@ihep.ac.cn}
\affiliation{Institute of High Energy Physics,
Chinese Academy of Sciences, Beijing 100049, China}
\affiliation{School of Physical Sciences, University of Chinese Academy of Sciences, Beijing 100049, China}

\date{27 November 2019}

\begin{abstract}
We present an updated analysis of the
coherent neutrino-nucleus elastic scattering data
of the COHERENT experiment
taking into account the new quenching factor
published recently in Phys. Rev. D100, 033003 (2019).
Through a fit of the COHERENT time-integrated energy spectrum,
we show that the new quenching factor
leads to a better determination of the average rms radius of the neutron distributions
of $^{133}\text{Cs}$ and $^{127}\text{I}$, while in combination with the atomic parity violation (APV) experimental results it allows to determine a data-driven APV measurement of the low-energy weak mixing angle in very good agreement with the Standard Model prediction.
We also find a
$3.7\sigma$
evidence of the suppression of coherence due to the nuclear structure.
Neutrino properties are better constrained by considering the
COHERENT time-dependent spectral data,
that allow us to improve the bounds on the neutrino charge radii
and magnetic moments.
We also present for the first time constraints on
the neutrino charges obtained with
coherent neutrino-nucleus elastic scattering data.
In particular, we obtain the first laboratory constraints on
the diagonal charge of $\nu_{\mu}$ and
the $\nu_{\mu}$-$\nu_{\tau}$ transition charge.
\end{abstract}


\maketitle

\section{Introduction}
\label{sec:introduction}

\bluechange{Coherent}
elastic neutrino-nucleus scattering is a new powerful tool
that allows to probe neutrino, electroweak and nuclear physics,
after its first, and so far only, observation in the COHERENT experiment~\cite{Akimov:2017ade}.
This process was predicted a long time ago~\cite{Freedman:1973yd,Freedman:1977xn,Drukier:1983gj},
but it eluded experimental detection because of the difficulty to observe
nuclear recoils with a very small kinetic energy $T$ of a few keV.
This is necessary for the coherent recoil of the nucleus
which occurs for 
$|\vec{q}| R \ll 1$~\cite{Bednyakov:2018mjd},
where $|\vec{q}| \simeq \sqrt{2 M T}$ is the three-momentum transfer,
$R$ is the nuclear radius of a few fm,
and
$M$ is the nuclear mass,
of the order of 100 GeV for heavy nuclei.

The measurements of the COHERENT experiment produced interesting results for
nuclear physics~\cite{Cadeddu:2017etk,Papoulias:2019lfi},
neutrino properties and interactions~\cite{Coloma:2017ncl,Liao:2017uzy,Kosmas:2017tsq,Denton:2018xmq,AristizabalSierra:2018eqm,Cadeddu:2018dux},
weak interactions~\cite{Cadeddu:2018izq,Huang:2019ene}, and
physics beyond the Standard Model~\cite{Dutta:2019eml,Dutta:2019nbn}.
One of the limiting factors of these analyses was the poor knowledge of
the quenching factor $f_{\text{Q}}(T)$ of the COHERENT CsI detector,
that is the ratio between the scintillation light emitted in nuclear and electron recoils
and determines the relation between the number of detected photoelectrons
$N_{\text{PE}}$
and the nuclear recoil kinetic energy $T$:
\begin{equation}
N_{\text{PE}}
=
\eta \, f_{\text{Q}}(T) \, Y_{\text{L}} \, T
,
\label{qfc}
\end{equation}
where
$ Y_{\text{L}} = 13.35 \, N_{\text{PE}} / \text{keV} $
is the light yield of the phototubes
and $\eta$ is a normalization factor.
In the original COHERENT publication~\cite{Akimov:2017ade}
the quenching factor $f_{\text{Q}}(T)$
was estimated to be constant with value $0.0878 \pm 0.0166$
between about 5 and 30 detected photoelectrons,
which correspond to $T$ from 4.3 to 25.6 keV.
The recent new accurate measurement of the quenching factor in Ref.~\cite{Collar:2019ihs}
reduced the relative uncertainty of the quenching factor from
$18.9\%$ to $5.1\%$
and, together with other revisited previous measurements,
provided the behavior of $f_{\text{Q}}(T)$ as a function of $T$,
improving the constant approximation in Ref.~\cite{Akimov:2017ade}\footnote{
The quenching factor in Ref.~\cite{Collar:2019ihs} is, however,
not supported by the COHERENT collaboration
[private communication received after the completion of this work].
}.
This significant refinement solicits a revision of the
results for neutrino, electroweak and nuclear physics
obtained from the analysis of the COHERENT data.
During the completion of this work,
two analyses of this type appeared on arXiv~\cite{Papoulias:2019txv,Khan:2019mju}.
Here we present the results of our analysis,
which has some differences in the method and results.
In particular, as we emphasized in Ref.~\cite{Cadeddu:2018dux},
the arrival time information of the COHERENT data~\cite{Akimov:2018vzs},
that was not considered in Refs.~\cite{Papoulias:2019txv,Khan:2019mju},
is important for distinguishing between the properties and interactions
of $\nu_{e}$ and $\nu_{\mu}$,
that are produced in the Oak Ridge Spallation Neutron Source
by different processes:
$\nu_\mu$'s are produced
from $\pi^+$ decays at rest
($\pi^+\to \mu^++\nu_\mu$)
and arrive at the COHERENT detector as a prompt signal within about
$1.5 \, \mu\text{s}$
after protons-on-targets;
$\bar\nu_{\mu}$'s and $\nu_e$'s are produced by $\mu^{+}$ decays at rest
($\mu^{+} \to e^{+} + \nu_{e} + \bar\nu_{\mu}$)
and arrive at the detector in a relatively longer time interval of about
$10 \, \mu\text{s}$.
In Ref.~\cite{Cadeddu:2018dux} we have shown that the
analysis of the time-dependent COHERENT spectrum
allows one to improve the constraints on the neutrino charge radii.
Here, we present in Section~\ref{sec:radii} an update of that analysis taking into account
the new quenching factor
and correcting the treatment of the sign of the contributions of the antineutrino charge radii (see the discussion in Section~\ref{sec:radii}).
We also present in Section~\ref{sec:charges} a new analysis
of the COHERENT data that allows to constrain the neutrino charges
(sometimes called millicharges because of their smallness).
In particular, we obtain the first laboratory constraints on
the diagonal charge of $\nu_{\mu}$ and
the $\nu_{\mu}$-$\nu_{\tau}$ transition charge.

The plan of the paper is as follows.
In Sections~\ref{sec:neutron}, \ref{sec:electroweak}, and \ref{sec:radii}
we update, respectively, the results on
the average rms radius of the neutron distributions in CsI,
on the weak mixing angle
and on the neutrino charge radii presented in Refs.~\cite{Cadeddu:2017etk,Cadeddu:2018izq,Cadeddu:2018dux},
taking into account the new quenching factor in the COHERENT experiment and, in the case of the weak mixing angle, also a new determination of the vector transition polarizability~\cite{Toh:2019iro}.
In Sections~\ref{sec:charges} and \ref{sec:magnetic}
we present, respectively, new constraints on the neutrino electric charges and magnetic moments.
Finally, in Section~\ref{sec:conclusions} we summarize the results of the paper.

\section{Radius of the nuclear neutron distribution}
\label{sec:neutron}

\bluechange{
The observation of coherent elastic neutrino-nucleus scattering
can be used to probe the nuclear neutron distribution~\cite{Patton:2012jr,Cadeddu:2017etk,Papoulias:2019lfi,Ciuffoli:2018qem,Papoulias:2019lfi}.
}
The standard weak-interaction differential cross section
for coherent elastic scattering of a neutrino with energy $E$
and a spin-zero nucleus $\mathcal{N}$
with $Z$ protons and $N$ neutrons is given by
\begin{equation}
\dfrac{d\sigma_{\nu_{\ell}\text{-}\mathcal{N}}}{d T}
(E,T)
=
\dfrac{G_{\text{F}}^2 M}{\pi}
\left(
1 - \dfrac{M T}{2 E^2}
\right)
\left[
g_{V}^{p}
Z
F_{Z}(|\vet{q}|^2)
+
g_{V}^{n}
N
F_{N}(|\vet{q}|^2)
\right]^2
,
\label{cs-std}
\end{equation}
where $G_{\text{F}}$ is the Fermi constant, $\ell = e, \mu, \tau$ is the neutrino flavour and
\begin{equation}
g_{V}^{p}
=
\dfrac{1}{2} - 2 \sin^2\!\vartheta_{W}
,
\qquad
g_{V}^{n}
=
- \dfrac{1}{2}
,
\label{gV}
\end{equation}
where $\vartheta_{\text{W}}$ is the weak mixing angle, also known as the Weinberg angle.
In Eq.~(\ref{cs-std})
$F_{Z}(|\vet{q}|^2)$
and
$F_{N}(|\vet{q}|^2)$
are, respectively, the form factors of the proton and neutron distributions in the nucleus.
They are given by the Fourier transform of the corresponding nucleon
distribution in the nucleus and
describe the loss of coherence for
$|\vet{q}| R_{p} \gtrsim 1$
and
$|\vet{q}| R_{n} \gtrsim 1$,
where $R_{p}$ and $R_{n}$ are, respectively, the rms radii of the proton and neutron distributions.
Since different parameterizations of the form factors are practically equivalent in the
analysis of COHERENT data~\cite{Cadeddu:2017etk},
we consider only the Helm parameterization~\cite{Helm:1956zz}
\begin{equation}
F(|\vet{q}|^2)
=
3
\,
\dfrac{j_{1}(|\vet{q}| R_{0})}{|\vet{q}| R_{0}}
\,
e^{- |\vet{q}|^2 s^2 / 2}
,
\label{ffHelm}
\end{equation}
where
$
j_{1}(x) = \sin(x) / x^2 - \cos(x) / x
$
is the spherical Bessel function of order one,
$s = 0.9 \, \text{fm}$ \cite{Friedrich:1982esq}
is the surface thickness
and $R_{0}$ is related to the rms radius $R$ by
$ R^2 = 3 R_{0}^2 / 5 + 3 s^2 $.
For the rms radii of the proton distributions of
$^{133}\text{Cs}$ and $^{127}\text{I}$
we adopt the values determined with high accuracy from
muonic atom spectroscopy~\cite{Angeli:2013epw}:
\begin{equation}
R_{p}({}^{133}\text{Cs}) = 4.8041 \pm 0.0046 \, \text{fm}
,
\qquad
R_{p}({}^{127}\text{I}) = 4.7500 \pm 0.0081 \, \text{fm}
.
\label{Rp}
\end{equation}

\begin{figure}[!t]
\centering
\includegraphics*[width=\linewidth]{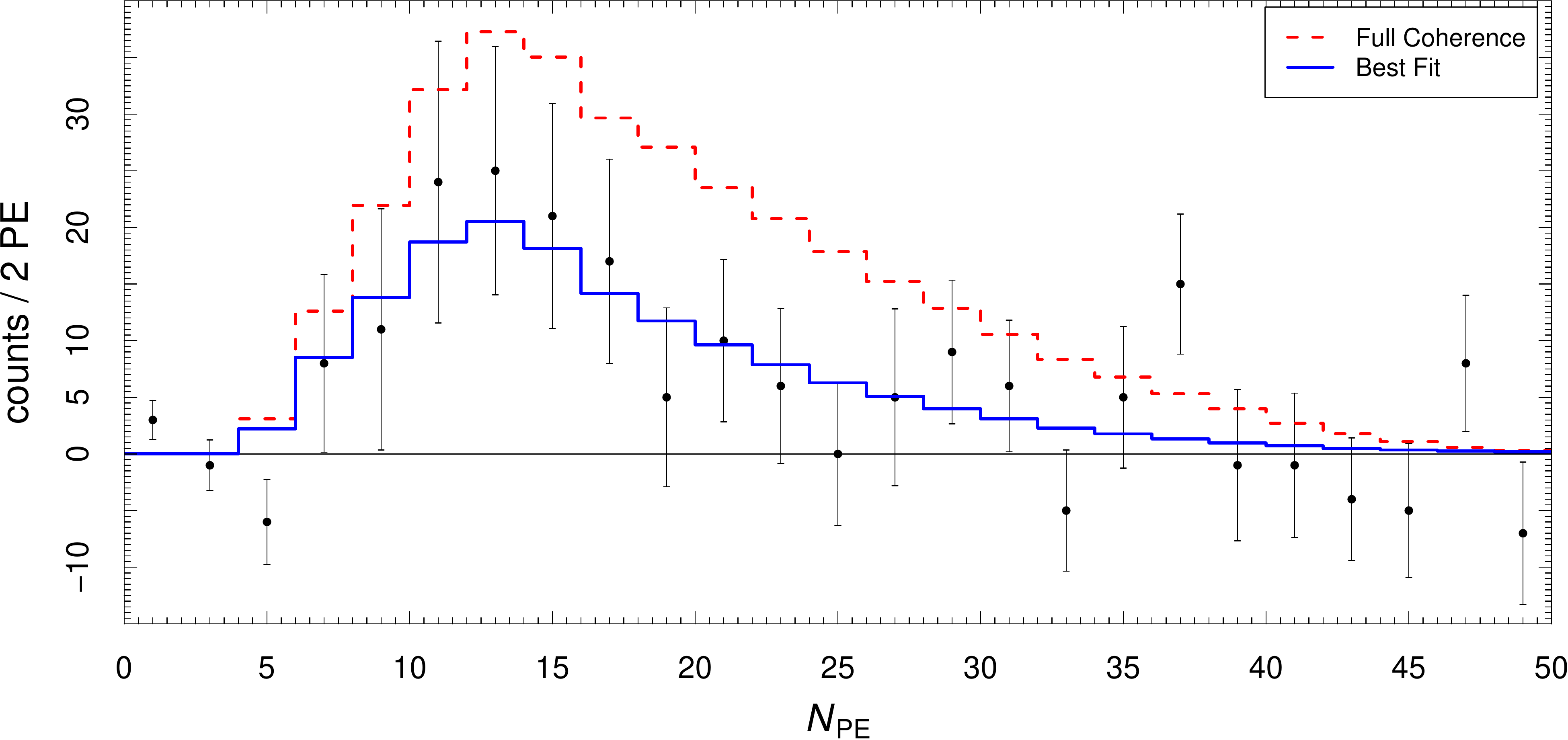}
\caption{ \label{fig:hist}
Histograms representing the fits of the COHERENT data \cite{Akimov:2017ade}
(black points with errorbars)
in the case of full coherence (red dashed)
and with the best-fit neutron distribution form factor
(blue solid).
}
\end{figure}

We fitted the COHERENT data in order to determine the average neutron rms radius
$R_{n}$
of $^{133}\text{Cs}$ and $^{127}\text{I}$
by improving the analysis in Ref.~\cite{Cadeddu:2017etk}
taking into account the new quenching function in Eq.~(\ref{qfc}).
We considered the least-squares function
\begin{equation}
\chi^2_{\text{C}}
=
\sum_{i=4}^{15}
\left(
\dfrac{
N_{i}^{\text{exp}}
-
\left(1+\alpha_{\text{c}}\right) N_{i}^{\text{th}}
-
\left(1+\beta_{\text{c}}\right) B_{i}
}{ \sigma_{i} }
\right)^2
+
\left( \dfrac{\alpha_{\text{c}}}{\sigma_{\alpha_{\text{c}}}} \right)^2
+
\left( \dfrac{\beta_{\text{c}}}{\sigma_{\beta_{\text{c}}}} \right)^2
+
\left( \dfrac{\eta-1}{\sigma_{\eta}} \right)^2
.
\label{chi-spectrum}
\end{equation}
For each energy bin $i$,
$N_{i}^{\text{exp}}$ is the experimental event number,
$N_{i}^{\text{th}}$
is the theoretical event number
that is calculated as explained in Refs.~\cite{Cadeddu:2017etk,Cadeddu:2018dux},
$B_{i}$ is the estimated number of background events, and
$\sigma_{i}$ is the statistical uncertainty.
We considered only the 12 energy bins from $i=4$ to $i=15$
of the COHERENT spectrum, because they cover the recoil kinetic energy of the new
Chicago-3 quenching factor measurement~\cite{Collar:2019ihs},
where the value of the quenching factor and its uncertainties are more reliable.
In Eq.~(\ref{chi-spectrum}),
$\alpha_{\text{c}}$ and $\beta_{\text{c}}$
are nuisance parameters which quantify,
respectively,
the systematic uncertainty of the signal rate
and
the systematic uncertainty of the background rate,
with
corresponding standard deviations
$\sigma_{\alpha_{\text{c}}} = 0.112$
and
$\sigma_{\beta_{\text{c}}} = 0.25$
\cite{Akimov:2017ade}.
The value of $\sigma_{\alpha_{\text{c}}}$ is smaller than that considered in previous analyses
because the previous value (0.28) included the quenching factor uncertainty,
that in Eq.~(\ref{chi-spectrum}) is taken into account
through the factor $\eta$ in Eq.~(\ref{qfc}),
with
$\sigma_{\eta} = 0.051$
according to the new determination in Ref.~\cite{Collar:2019ihs}.
We calculated the new value of $\sigma_{\alpha_{\text{c}}}$
by summing in quadrature
the 5\% signal acceptance uncertainty
and
the 10\% neutron flux uncertainty
estimated by the COHERENT collaboration \cite{Akimov:2017ade},
without considering an estimated 5\% neutron form factor uncertainty
because we obtain the neutron form factor
from the data.

The COHERENT spectral data are shown in Figure~\ref{fig:hist}
together with the best-fit histogram,
that corresponds to
$(\chi^2_{\text{C}})_{\text{min}} = 3.0$
with
$11$
degrees of freedom.

\begin{figure}[!t]
\centering
\includegraphics*[width=0.5\linewidth]{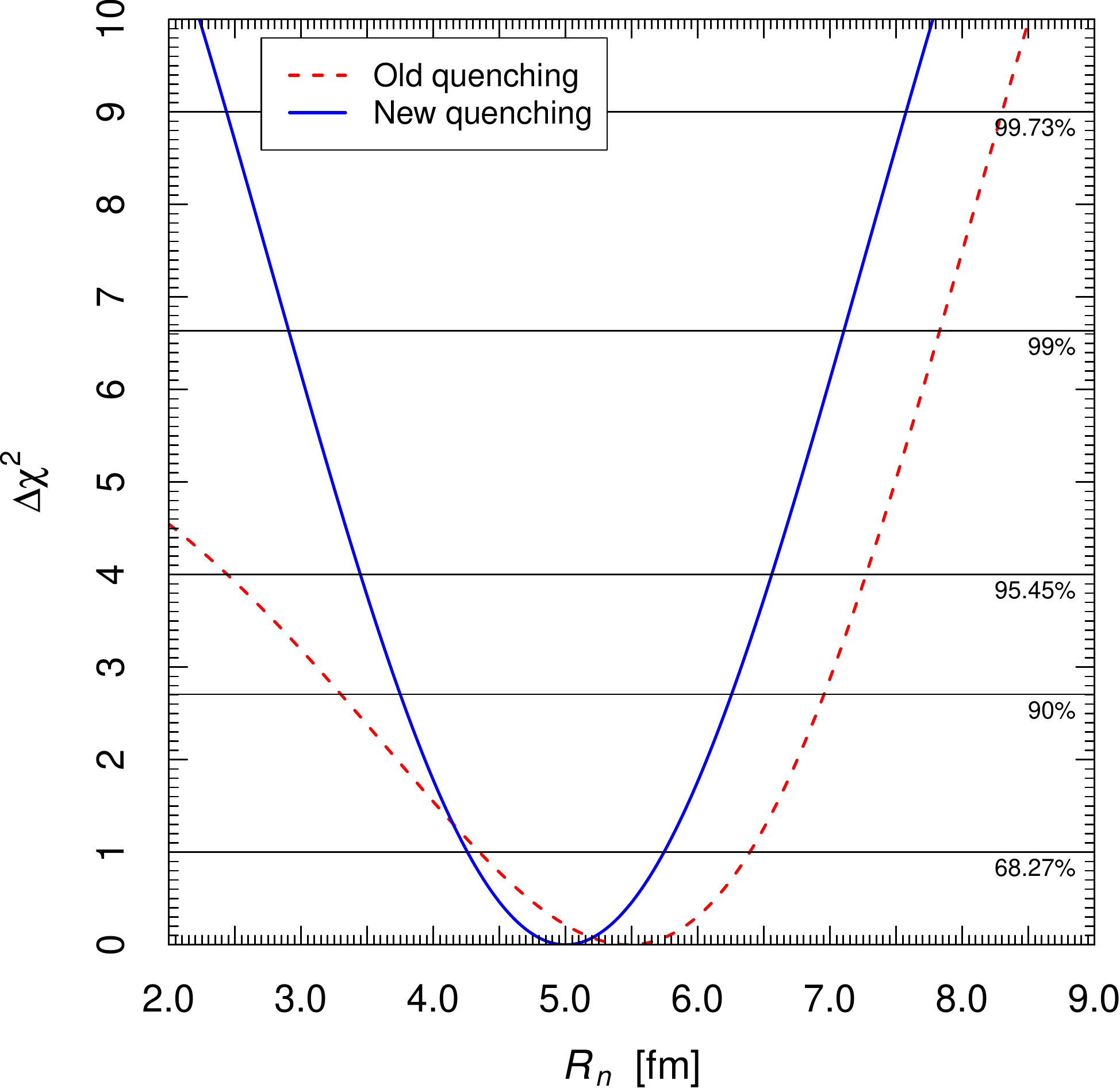}
\caption{ \label{fig:rn-chi}
$\Delta\chi^2 = \chi^2_{\text{C}} - (\chi^2_{\text{C}})_{\text{min}}$
as a function of the neutron rms radius $R_{n}$
obtained from the fit of the data of the COHERENT experiment
with the new quenching factor \protect\cite{Collar:2019ihs},
compared with the result in Ref.~\protect\cite{Cadeddu:2017etk}
obtained with the old quenching factor in Ref.~\protect\cite{Akimov:2017ade}.
}
\end{figure}

Figure~\ref{fig:hist} shows also the histogram obtained in the case of full coherence,
i.e. with the neutron and proton form factors set to unity in the cross section of Eq.~(\ref{cs-std}).
One can see that the fit is much worse than the best fit.
The corresponding $(\chi^2_{\text{C}})_{\text{min}}$ is
$18.1$
and the difference with that of the best fit implies a $p$-value of
$1.0
\times
10^{-4}$.
Therefore there is a
$3.7\sigma$
evidence of the suppression of coherence due to the nuclear structure.
This is an improvement with respect to the $2.3\sigma$ evidence found in Ref.~\cite{Cadeddu:2017etk}
with the old quenching factor~\cite{Akimov:2017ade}.

Figure~\ref{fig:rn-chi} shows the comparison of
$\Delta\chi^2 = \chi^2_{\text{C}} - (\chi^2_{\text{C}})_{\text{min}}$
as a function of the neutron rms radius $R_{n}$
obtained with the old~\cite{Akimov:2017ade}
and new~\cite{Collar:2019ihs}
quenching factors.
It is clear that the new quenching factor allows us to perform a better determination of $R_{n}$,
with smaller uncertainties, especially when considering large values of the confidence level.
In particular,
small values of $R_{n}$ are better constrained with the new quenching factor
and we obtain
\begin{equation}
R_{n}
=
5.0
{}^{+0.7}_{-0.7}
(1\sigma)
{}^{+1.5}_{-1.5}
(2\sigma)
{}^{+2.5}_{-2.6}
(3\sigma)
\, \text{fm}
.
\label{Rn-fit}
\end{equation}
It is interesting that the value of $R_{n}$ obtained with the new quenching factor
is smaller than that found in Ref.~\cite{Cadeddu:2017etk}
with the old quenching factor
($R_{n} = 5.5 {}^{+0.9}_{-1.1} \, \text{fm}$).
\bluechange{This shift goes in the direction of}
the theoretical nuclear model predictions,
that indicate a value around 5.0 fm (see Table~I in Ref.~\cite{Cadeddu:2017etk}).
For the neutron skin
$\Delta R_{np} = R_{n} - R_{p}$
we obtain
\begin{equation}
\Delta R_{np}
=
0.2
{}^{+0.7}_{-0.7}
(1\sigma)
{}^{+1.5}_{-1.5}
(2\sigma)
{}^{+2.5}_{-2.6}
(3\sigma)
\, \text{fm}
,
\label{Rnp-fit}
\end{equation}
that is in agreement with the theoretical nuclear model predictions
(see Figure~1 of Ref.~\cite{Horowitz:1999fk} and Table~I in Ref.~\cite{Cadeddu:2017etk}).

Our results have some differences with those in Refs.~\cite{Papoulias:2019txv,Khan:2019mju}.
The author of Ref.~\cite{Papoulias:2019txv}
found
$R_{n} = 5.1 {}^{+1.3}_{-1.5} \, \text{fm}$,
with a best fit similar to ours, but with larger uncertainties.
We suspect that this is due to the fact that the analysis of
Ref.~\cite{Papoulias:2019txv}
is based only on the total number of COHERENT events,
not on the COHERENT energy spectrum as ours.
The authors of Ref.~\cite{Khan:2019mju}
found
$R_{n} = 4.6 {}^{+0.9}_{-0.8} \, \text{fm}$,
where the uncertainties are only slightly larger than ours,
but the best fit is significantly lower.
We suspect that this is due to the fact that,
although the analysis of Ref.~\cite{Khan:2019mju}
is based on the fit of the COHERENT energy spectrum as ours,
the quenching factor was assumed to be constant.
This is a rather rough approximation,
as one can see from Figure~1 of Ref.~\cite{Collar:2019ihs}.

\section{Weak mixing angle}
\label{sec:electroweak}

A determination of the radius of the neutron distribution of $^{133}\text{Cs}$
as that in Eq.~(\ref{Rn-fit})
can be used to improve the evaluation of the low-energy value of the weak mixing angle $\vartheta_{\text{W}}$
obtained from the measurement of the weak charge $Q^{\text{Cs}}_W$ of $^{133}\text{Cs}$ in atomic parity violation (APV) experiments, also known as parity nonconservation (PNC) experiments.
In this Section we update the evaluation of $\vartheta_{\text{W}}$ obtained in
Ref.~\cite{Cadeddu:2018izq}
from the analysis of the COHERENT data
taking into account the new quenching factor.

In the SM the weak charge including electroweak corrections is related to the weak mixing angle through the relation~\cite{Erler:2013xha}
\begin{align}
\label{QWSMEWcorrections1}
Q_{{W}}^{\rm SM+rad. corr.} 
&\equiv - 2 [ Z (g_{AV}^{\, e p} + 0.00005) \\\nonumber
 &+ N (g_{AV}^{\, e n} + 0.00006) ] 
 \left( 1 - \dfrac{\alpha}{2 \pi} \right) \\\nonumber
&\approx Z (1 - 4 \sin^2 \theta_W^{\rm SM}) - N, 
\end{align}
where $\alpha$ is the fine-structure constant and the couplings of electrons to nucleons, $g_{AV}^{\, e p}$ and $g_{AV}^{\, e n}$, are given by
\begin{equation}
g_{AV}^{\, e p}  \approx - \dfrac{1}{2} + 2 \sin^2\theta_W^{\rm SM}, \,\,\,\,\mathrm{and} \,\,\,\, g_{AV}^{\, e n}  \approx \dfrac{1}{2}\,.
\end{equation}

\begin{figure}[!t]
\centering
\includegraphics*[width=0.7\linewidth]{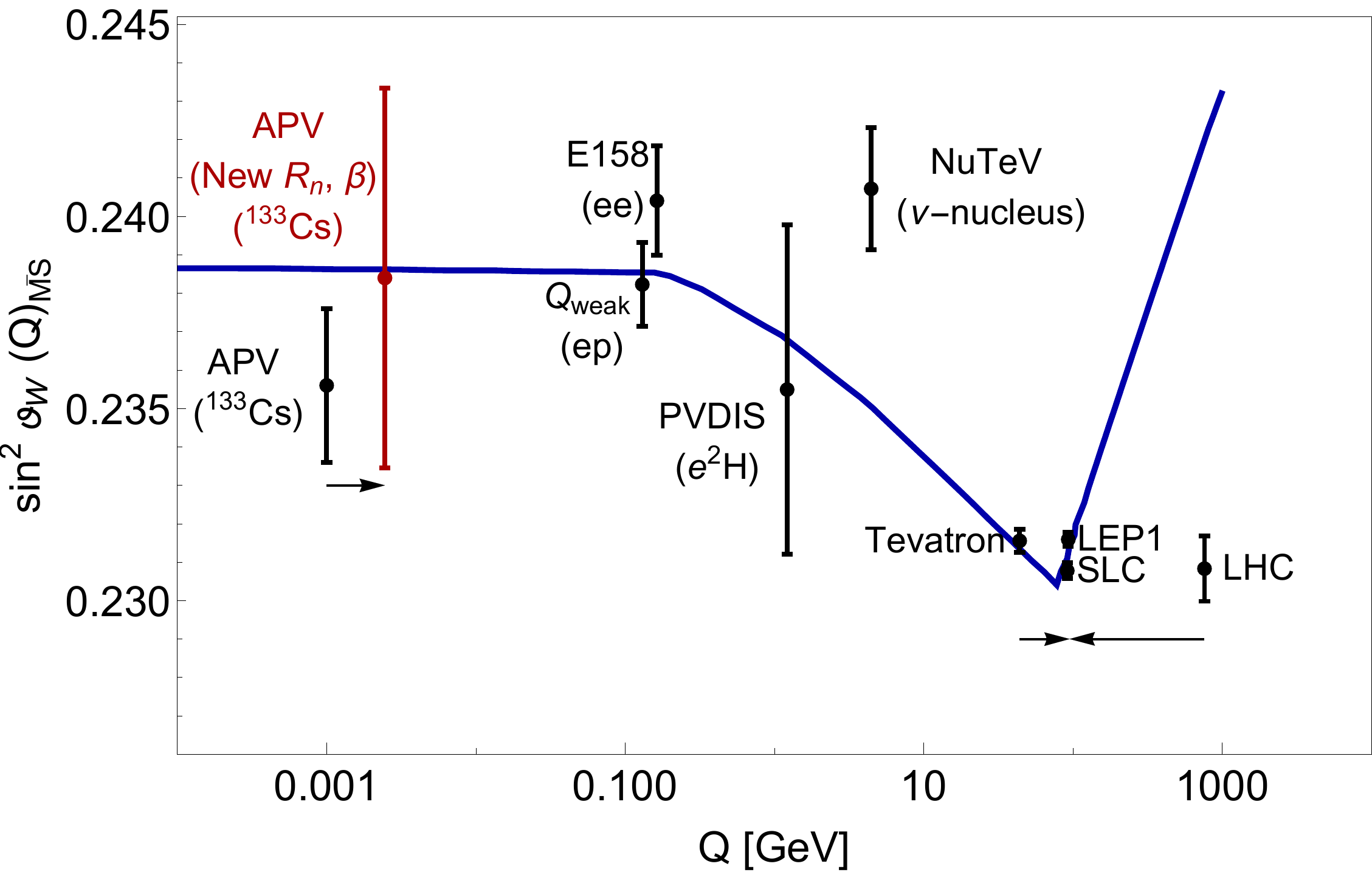}
\caption{ \label{fig:running}
Variation of $\sin^2 \vartheta_{\text{W}}$ with energy scale Q. The SM prediction is shown as the solid curve, together with
experimental determinations in black at the $Z$-pole~\cite{Tanabashi:2018oca} (Tevatron, LEP1, SLC, LHC),
from APV on Caesium~\cite{Wood:1997zq,Dzuba:2012kx}, which has a typical momentum transfer given by $\langle Q\rangle\simeq$~2.4 MeV, M{\o}ller scattering~\cite{Anthony:2005pm} (E158), deep inelastic scattering of polarized electrons on deuterons~\cite{Wang:2014bba} ($ e^2H $ PVDIS) and from
neutrino-nucleus scattering~\cite{Zeller:2001hh} (NuTeV) and the new result from the proton's weak charge 
at $Q = 0.158$ GeV~\cite{Androic:2018kni} ($ \text{Q}_{\text{weak}} $). In red it is shown the result derived in this paper, obtained correcting the APV data point by the improved direct Caesium neutron rms radius determination obtained in this work. For clarity we displayed the old APV point to the left and the Tevatron and LHC points horizontally to the left and to the right, respectively.}
\end{figure}

Experimentally, the weak charge of a nucleus is extracted from the ratio of the parity violating amplitude, $E_{\text{PNC}}$, to the Stark vector transition polarizability, $\beta$, and by calculating theoretically $E_{\rm PNC}$ in terms of \QW, leading to
\begin{equation}
\label{QWeq}
Q_W= N \left( \dfrac{{\rm Im}\, E_{\rm PNC}}{\beta} \right)_{\rm exp.} 
\left( \dfrac{Q_W}{N\, {\rm Im}\, E_{\rm PNC}} \right)_{\rm th.} \beta_{\rm exp.+th.}\,,
\end{equation}
where $\beta_{\rm exp.+th.}$ and $(\mathrm{Im}\, E_{\rm PNC})_{\rm th.}$ are determined from atomic theory, and Im stands for imaginary part (see Ref.~\cite{Tanabashi:2018oca}).
In particular, we use 
$({\rm Im}\, E_{\rm PNC}/{\beta})_{\rm exp} = (- 3.0988\pm0.0109) \times 10^{-13} |e|/a_B^2$~\cite{Wood:1997zq}, where $a_B$ is the Bohr radius and $|e|$ is the electric charge. Differently from Ref.~\cite{Cadeddu:2018izq}, we use a more recent determination of the vector transition polarizability, namely $\beta_{\rm exp.+th.} = (27.139\pm0.042)\, a_B^3$~\cite{Toh:2019iro}, which has a smaller uncertainty and differs by almost 2$\sigma$ with respect to the previous determination of it, coming instead from a calculation of the hyperfine changing contribution to the magnetic dipole matrix element $M1_{hf}$~\cite{PhysRevA.62.052101}. For the imaginary part of $E_{\rm PNC}$ we use
$({\rm Im}\, E_{\rm PNC})_{\rm th.}=(0.8977\pm0.0040)\times10^{-11}|e|a_B \frac{Q_W}{N}$~\cite{Dzuba:2012kx}, which includes corrections for many-body effects that were neglected in previous works.
\bluechange{
The authors of Ref.~\cite{Dzuba:2012kx} introduced also a correction, called ``neutron skin", to take into account the difference between $R_n$ and $R_p$ that is not considered in the nominal atomic theory derivation. Here, we want to remove this correction in order to re-evaluate it using the direct measurement of $R_n$ shown in Eq.~(\ref{Rn-fit}).
Indeed, at the time of Ref.~\cite{Dzuba:2012kx}, this correction could only have been estimated from an extrapolation of data from antiprotonic atom x-ray. 
Removing the neutron skin correction from the total value of $({\rm Im}\, E_{\rm PNC})_{\rm th.}$ (taking the values from table IV of Ref.~\cite{Dzuba:2012kx}) the theoretical amplitude, referred to as ``without neutron skin" and indicated by the apex w.n.s., becomes}
\begin{equation}
\label{MyImEPNC}
({\rm Im}\, E_{\rm PNC})_{\rm th.}^{\rm w.n.s.}=(0.8995\pm0.0040)\times10^{-11}|e|a_B \frac{Q_W}{N}\,.
\end{equation}
\bluechange{
Differently from Ref.~\cite{Cadeddu:2018izq}, we use the neutron skin correction parameterization well summarized in Ref.~\cite{Viatkina_2019} to correct the value of $({\rm Im}\, E_{\rm PNC})_{\rm th.}^{\rm w.n.s.}$. Namely, the corrected value, indicated with the apex n.s., depends upon the value of $R_n$
\begin{equation}
(E_{\rm PNC})_{\rm th.}^{\rm n.s.}(R_n)=(E_{\rm PNC})_{\rm th.}^{\rm w.n.s.}+\delta E^\mathrm{n.s.}_\mathrm{PNC}(R_n)=(E_{\rm PNC})_{\rm th.}^{\rm w.n.s.}+\left[ \frac{\mathrm{N}}{Q_W^{\mathrm{SM+rad. corr.}}}\left(1-\frac{q_n(R_n)}{q_p}\right)\cdot E_\mathrm{PNC}^\mathrm{w.n.s.} \right] \,,
\label{eq:corr1}
\end{equation}
where $q_p$ and $q_n$ are factors which incorporate the radial dependence of the electron axial transition matrix element considering the proton and the neutron spatial distribution, respectively, as defined in Ref.~\cite{Cadeddu:2018izq}. 
A frequently used estimate of $q_n$ and $q_p$ is the one obtained~\cite{PhysRevC.46.2587,Pollock1999,Horowitz2001} assuming a uniform nuclear charge distribution and $R_n \approx R_p$, namely
\begin{eqnarray}
q_p &\approx& 1-(Z\alpha)^2(0.26)\\
q_n(R_n) &\approx& 1-(Z\alpha)^2\left(0.26+0.221 \left( \frac{R^2_n}{R^2_p}-1 \right)\right)\,. \nonumber
\end{eqnarray}
In particular, it is possible to see the dependence of $q_n$ on the values of $R_n$. 
The neutron skin corrected value of the weak charge depends thus on the value of $R_n$ and it can be written as
\begin{equation}
\label{QWeq_step1}
Q_W^\mathrm{n.s.}(R_n)= N \left( \frac{{\rm Im}\, E_{\rm PNC}}{\beta} \right)_{\rm exp.} 
\left( \frac{Q_W}{N\, {\rm Im}\,( E_{\rm PNC}^\mathrm{w.n.s.}+\delta E^\mathrm{n.s.}_\mathrm{PNC}(R_n))} \right)_{\rm th.} \beta_{\rm exp.+th.}\,,
\end{equation}

Using the value of $R_n$ in Eq.~(\ref{Rn-fit}) found with the new quenching factor, we derive the updated data-driven correction to the $E_{\rm PNC}$ theoretical amplitude, namely $\delta E^\mathrm{n.s.}_\mathrm{PNC}(R_n=5.0\, \mathrm{fm})=-0.0030\times10^{-11}i|e|a_B \frac{Q_W}{N}$.
Thanks to this data-driven correction, we determine an updated experimental value of the weak charge in Cs, which takes into account the difference between the measured values of $R_n$ and $R_p$, that can be directly compared with the SM prediction $Q_W^{\rm SM+rad. corr.}=-73.23\pm0.01$~\cite{Tanabashi:2018oca}, namely}
\begin{equation}
Q_W^{\rm Cs\,n.s.}(R_n=5.0\, \mathrm{fm}) = -73.2\pm1.1,
\end{equation}
where the final uncertainty is at 1$\sigma$. 
Our result on the weak charge of $^{133}\text{Cs}$ is different and with a much reduced uncertainty with respect to
that obtained recently in Ref.~\cite{Khan:2019mju}
with the new quenching factor.
Besides the reasons mentioned at the end of Section~\ref{sec:neutron}, there is also the usage of a more recent determination of the vector transition polarizability.\\
The value of $Q_W^{\rm Cs}$ obtained in this way relies on the improved direct experimental input for $R_n$ of $^{133}\text{Cs}$, and allows to determine the new APV value of the weak mixing angle
\begin{equation}
\sin^2 \vartheta_{\text{W}}=0.238\pm0.005,
\end{equation}
with a central value in very good agreement with the SM at low momentum transfer, as shown by the red point in Figure~\ref{fig:running}\bluechange{, where a summary of the weak mixing angle measurements as a function of the energy scale $Q$ is shown along with the SM predicted running of $\sin^2 \vartheta_{\text{W}}$, calculated in the so-called modified minimal subtraction ($\overline{\text{MS}}$) renormalization scheme~\cite{Tanabashi:2018oca, Erler:2004in,Erler:2017knj}}.

Following the approach developed in Ref.~\cite{Cadeddu:2018izq}, the APV data can be fitted simultaneously with the COHERENT data to determine a combined value of $R_n$ for $^{133}\text{Cs}$.
Assuming the PDG value~\cite{Tanabashi:2018oca} of the weak mixing angle at low momentum transfer\footnote{Note that this procedure is perfectly consistent since the neutron radius from COHERENT data has been obtained assuming the same low-energy value of the weak mixing angle.}, the following combined APV and COHERENT least-squares function can be built
\begin{align}
\chi^2
=
\null & \null
\chi^2_{\text{C}}+\chi_{\textrm{APV}}^2
\nonumber
\\
=
\null & \null
\chi^2_{\text{C}}
+
\left(
\dfrac{
(Q_W^{\rm Cs\,n.s.}(R_n))
-
Q_W^{\rm SM+rad. corr.}
}{ \sigma_{APV} }
\right)^2
\,,
\label{chitot}
\end{align} 
where the first term is defined in Eq.~(\ref{chi-spectrum}) and the second term represents the least-squares function corresponding to the APV measurement for $^{133}\text{Cs}$,
in which $\sigma_{APV}$ is the total uncertainty corresponding to 0.43. 
 
\begin{figure}[!t]
\centering
\includegraphics*[width=0.5\linewidth]{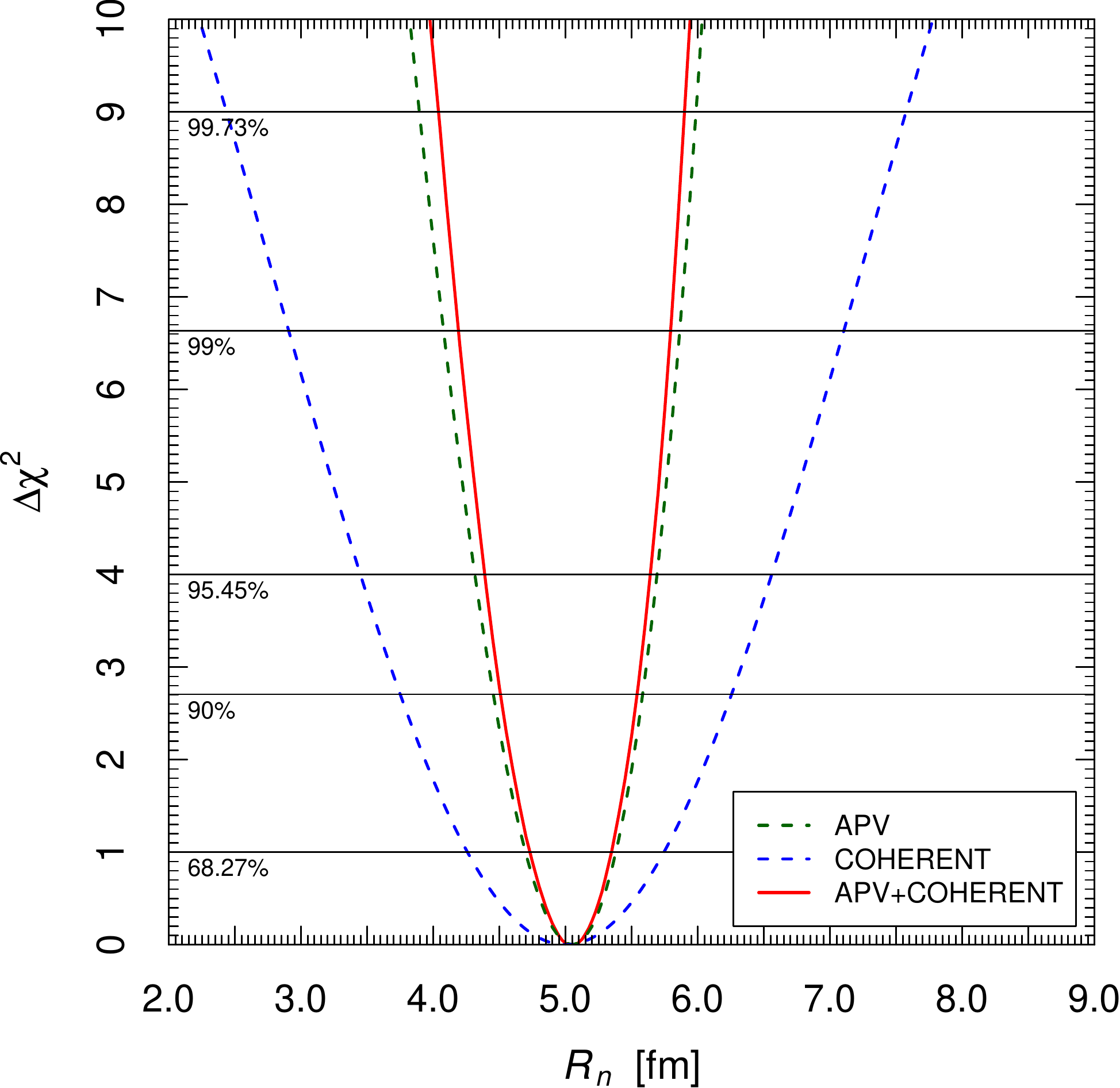}
\caption{ \label{fig:deltachi2} The solid red line shows the $\Delta \chi^2=\chi^2-\chi_{\rm min}^2$, with $\chi^2$ as defined in Eq.~(\ref{chitot}), as a function of the neutron rms radius, $R_{n}$, obtained from the combined fit of the COHERENT data, with the new quenching factor~\protect\cite{Collar:2019ihs}, and the APV Caesium measurement, with a new determination of the vector transition polarizability~\cite{Toh:2019iro}. The blue and the green dashed lines show the contribution of the separate fits of the COHERENT and APV datasets, respectively.
} 
\end{figure}
The solid red line in Figure~\ref{fig:deltachi2} shows the corresponding marginal values of the $\chi^2$
as a function of $R_{n}$, while for completeness the results of the fit to the COHERENT and APV dataset alone is shown separately by the blue and green dashed lines, respectively. 
Thanks to the usage of the more recent determination of the vector transition polarizability~\cite{Toh:2019iro}, the two dataset point to a similar value of $R_n$. One can see that the inclusion of the APV measurement allows to shrink significantly the $\Delta\chi^2$ profile, reducing by more than half the uncertainty. The result of the combined APV and COHERENT measurement is
\begin{equation}
R_{n}=5.04\pm 0.31\,\text{fm},
\label{rn}
\end{equation}. Using the value found in Eq.~(\ref{rn}) and the updated value of $R_{p}$ of $^{133}\text{Cs}$ in Eq.~(\ref{Rp}), it is possible to infer a more precise value of the $^{133}\text{Cs}$ neutron skin, which is 
\begin{equation}
\Delta R_{np}=0.23\pm 0.31\,\text{fm},
\label{nskin}
\end{equation}
in excellent agreement with the model-predicted values~\cite{Horowitz:1999fk,Cadeddu:2017etk}.

\section{Neutrino charge radii}
\label{sec:radii}

In the Standard Model of electroweak interactions
neutrinos are exactly neutral particles,
but they have the charge radii induced by radiative corrections given by
\cite{Bernabeu:2000hf,Bernabeu:2002nw,Bernabeu:2002pd}
\begin{equation}
\langle{r}_{\nu_{\ell}}^{2}\rangle_{\text{SM}}
=
-
\frac{G_{\text{F}}}{2\sqrt{2}\pi^{2}}
\left[
3-2\ln\left(\frac{m_{\ell}^{2}}{m^{2}_{W}}\right)
\right]
,
\label{G050}
\end{equation}
where $m_{W}$ and $m_{\ell}$ are the $W$ boson and charged lepton masses
and we use the conventions in Refs.~\cite{Giunti:2014ixa,Cadeddu:2018dux}.
Note that the Standard Model charge radii are diagonal in the flavor basis,
because the generation lepton numbers are conserved.
Numerically, we have
\begin{align}
\null & \null
\langle{r}_{\nu_{e}}^{2}\rangle_{\text{SM}}
=
- 0.83 \times 10^{-32} \, \text{cm}^{2}
,
\label{reSM}
\\
\null & \null
\langle{r}_{\nu_{\mu}}^{2}\rangle_{\text{SM}}
=
- 0.48 \times 10^{-32} \, \text{cm}^{2}
,
\label{rmSM}
\\
\null & \null
\langle{r}_{\nu_{\tau}}^{2}\rangle_{\text{SM}}
=
- 0.30 \times 10^{-32} \, \text{cm}^{2}
.
\label{rtSM}
\end{align}
Since the current 90\% CL experimental bounds for
$\langle{r}_{\nu_{e}}^{2}\rangle$
and
$\langle{r}_{\nu_{\mu}}^{2}\rangle$
listed in Table~I of Ref.~\cite{Cadeddu:2018dux}
are about one order of magnitude larger than the Standard Model predictions,
there are good hopes that these values can be probed in a near future.

In Ref.~\cite{Cadeddu:2018dux} we have shown that
the COHERENT elastic neutrino-nucleus scattering data
allows to constrain
not only the flavor-diagonal neutrino charge radii,
but also the transition charge radii\footnote{\bluechange{
In theories beyond the Standard Model
neutrinos can have flavor transition charge radii
$\langle{r}_{\nu_{\ell\ell'}}^{2}\rangle$
with $\ell\neq\ell'$
(see the review in Ref.~\cite{Giunti:2014ixa}).
As explained in Ref.~\cite{Cadeddu:2018dux},
even if the matrix of the neutrino charge radii is diagonal in the mass basis,
transition charge radii in the flavor basis are generated by neutrino mixing.
}},
taking into account their contribution discussed for the first time in Ref.~\cite{Kouzakov:2017hbc}
in the context of elastic neutrino-electron scattering.
Here we present the update of that analysis taking into account the new quenching factor.
We also correct an unfortunate mistake introduced in Ref.~\cite{Cadeddu:2018dux}
in the treatment of the sign of the contributions of the antineutrino charge radii
by noting that neutrinos and antineutrinos have opposite charge radii.
This is correct,
but it was not noted that also the weak neutral current couplings
change sign from neutrinos to antineutrinos.
This is due to the fact that both
the electromagnetic vector current
and
the $V-A$ weak neutral-current
change sign under a CP transformation
that changes left-handed neutrinos into right-handed antineutrinos.
Therefore the relative sign
of the weak neutral current and charge radius contributions
does not change from neutrinos to antineutrinos.

For simplicity, we consider only two of the cases
discussed in Ref.~\cite{Cadeddu:2018dux},
i.e. the fits of the COHERENT
time-dependent energy spectrum with fixed and free values
of the rms radii of the neutron distributions of
$^{133}\text{Cs}$ and $^{127}\text{I}$.
In the first case we \bluechange{consider} the same values assumed in Ref.~\cite{Cadeddu:2018dux}:
\begin{equation}
R_{n}({}^{133}\text{Cs}) = 5.01 \, \text{fm}
,
\qquad
R_{n}({}^{127}\text{I}) = 4.94 \, \text{fm}
,
\label{RnRMF}
\end{equation}
which have been obtained in the relativistic mean field (RMF) NL-Z2 \cite{Bender:1999yt}
nuclear model calculated in Ref.~\cite{Cadeddu:2017etk}.

As we emphasized in Ref.~\cite{Cadeddu:2018dux} and in the introduction,
the arrival time information of the COHERENT data~\cite{Akimov:2018vzs}
is important for distinguishing between the properties
of $\nu_{e}$ and $\nu_{\mu}$
and in particular for the determination of the charge radii.
Therefore,
in this case,
instead of the least-squares function in Eq.~(\ref{chi-spectrum}),
we consider the Poissonian least-squares function~\cite{Baker:1983tu}
\begin{align}
\chi^2
=
\null & \null
2
\sum_{i=4}^{15}
\sum_{j=1}^{12}
\left[
\left( 1 + \alpha_{\text{c}} \right) N_{ij}^{\text{th}}
+
\left( 1 + \beta_{\text{c}} \right) B_{ij}
+
\left( 1 + \gamma \right) N_{ij}^{\text{bck}}
-
N_{ij}^{\text{C}}
+
N_{ij}^{\text{C}}
\ln\!\left(
\frac{ N_{ij}^{\text{C}} }{
\left( 1 + \alpha_{\text{c}} \right) N_{ij}^{\text{th}}
+
\left( 1 + \beta_{\text{c}} \right) B_{ij}
+
\left( 1 + \gamma \right) N_{ij}^{\text{bck}}
}
\right)
\right]
\nonumber
\\
\null & \null
\hspace{2cm}
+
\left( \frac{\alpha_{\text{c}}}{\sigma_{\alpha_{\text{c}}}} \right)^2
+
\left( \frac{\beta_{\text{c}}}{\sigma_{\beta_{\text{c}}}} \right)^2
+
\left( \frac{\gamma}{\sigma_{\gamma}} \right)^2
+
\left( \dfrac{\eta-1}{\sigma_{\eta}} \right)^2
,
\label{chi2time}
\end{align}
that allows us to consider time-energy bins
with few or zero events.
In Eq.~(\ref{chi2time}),
$i$ is the index of the energy bins,
$j$ is the index of the time bins,
$N_{ij}^{\text{th}}$ are the theoretical predictions that depend on the neutrino charge radii,
$N_{ij}^{\text{C}}$ are the coincidence (C) data, which contain signal and background events,
$B_{ij}$ are the estimated neutron-induced backgrounds, and
$N_{ij}^{\text{bck}}$ are the estimated backgrounds obtained from the anti-coincidence (AC) data
given in the COHERENT data release~\cite{Akimov:2018vzs}.
The parameters $\alpha_{\text{c}}$, $\beta_{\text{c}}$, and $\eta$ are the same as in the least-square function in Eq.~(\ref{chi-spectrum})
that we used in the analysis of the time-integrated COHERENT data.
The nuisance parameter $\gamma$
and its uncertainty
$\sigma_{\gamma} = 0.05$ quantify the systematic uncertainty of the background estimated from the AC data.

The theoretical predictions $N_{ij}^{\text{th}}$ in Eq.~(\ref{chi2time})
have been calculated with the differential neutrino-nucleus
($\nu_{\ell}\text{-}\mathcal{N}$)
cross section
\begin{equation}
\dfrac{d\sigma_{\nu_{\ell}\text{-}\mathcal{N}}}{d T}
(E,T)
=
\dfrac{G_{\text{F}}^2 M}{\pi}
\left(
1 - \dfrac{M T}{2 E^2}
\right)
\left\{
\left[
\left( g_{V}^{p} - \tilde{Q}_{\ell\ell} \right)
Z
F_{Z}(|\vet{q}|^2)
+
g_{V}^{n}
N
F_{N}(|\vet{q}|^2)
\right]^2
+
Z^2
F_{Z}^2(|\vet{q}|^2)
\sum_{\ell'\neq\ell}
|\tilde{Q}_{\ell'\ell}|^2
\right\}
,
\label{cs-chr}
\end{equation}
with the contributions of the charge radii
$\langle{r}_{\nu_{\ell\ell'}}^2\rangle$
in the flavor basis expressed through~\cite{Kouzakov:2017hbc}
\begin{equation}
\tilde{Q}_{\ell\ell'}
=
\frac{2}{3} \, m_{W}^2 \sin^2\!\vartheta_{W} \langle{r}_{\nu_{\ell\ell'}}^2\rangle
=
\dfrac{ \sqrt{2} \pi \alpha }{ 3 G_{\text{F}} }
\, \langle{r}_{\nu_{\ell\ell'}}^2\rangle
.
\label{Qchr}
\end{equation}
In the case of
$\bar\nu_{\ell}\text{-}\mathcal{N}$ scattering,
we have
$g_{V}^{p,n} \to - g_{V}^{p,n}$
and
$\langle{r}_{\nu_{\ell\ell'}}\rangle
\to
\langle{r}_{\bar\nu_{\ell\ell'}}\rangle = - \langle{r}_{\nu_{\ell\ell'}}\rangle$,
as explained above.
Hence,
the charge radii of neutrinos and antineutrinos
contribute with the same sign to the shift of
$\sin^2\!\vartheta_{W}$.

\begin{table*}[t!]
\begin{center}
\begin{tabular}{c|ccc|ccc|}
\\
&
\multicolumn{3}{c|}{Fixed $R_{n}$}
&
\multicolumn{3}{c|}{Free $R_{n}$}
\\
&
Best Fit
&
90\% CL
&
99\% CL
&
Best Fit
&
90\% CL
&
99\% CL
\\
\hline
$\langle{r}_{\nu_{ee}}^2\rangle$
&
$-26$
&
$ -56 \div 7 $
&
$ -65 \div 16 $
&
$-25$
&
$ -56 \div 10 $
&
$ -65 \div 20 $
\\
$\langle{r}_{\nu_{\mu\mu}}^2\rangle$
&
$-30$
&
$ -60 \div 10 $
&
$ -64 \div 14 $
&
$-32$
&
$ -60 \div 14 $
&
$ -64 \div 17 $
\\
$|\langle{r}_{\nu_{e\mu}}^2\rangle|$
&
0
&
$ < 28 $
&
$ < 32 $
&
0
&
$ < 28 $
&
$ < 31 $
\\
$|\langle{r}_{\nu_{e\tau}}^2\rangle|$
&
0
&
$ < 32 $
&
$ < 40 $
&
0
&
$ < 32 $
&
$ < 41 $
\\
$|\langle{r}_{\nu_{\mu\tau}}^2\rangle|$
&
0
&
$ < 35 $
&
$ < 39 $
&
0
&
$ < 35 $
&
$ < 39 $
\\
\hline
$q_{\nu_{ee}}$
&
$14$
&
$ -8 \div 46 $
&
$ -18 \div 59 $
&
$16$
&
$ -9 \div 46 $
&
$ -20 \div 59 $
\\
$q_{\nu_{\mu\mu}}$
&
$-3$
&
$ -10 \div 18 $
&
$ -14 \div 34 $
&
$-2$
&
$ -11 \div 19 $
&
$ -16 \div 35 $
\\
$|q_{\nu_{e\mu}}|$
&
0
&
$ < 22 $
&
$ < 29 $
&
0
&
$ < 22 $
&
$ < 29 $
\\
$|q_{\nu_{e\tau}}|$
&
0
&
$ < 30 $
&
$ < 41 $
&
0
&
$ < 30 $
&
$ < 41 $
\\
$|q_{\nu_{\mu\tau}}|$
&
0
&
$ < 28 $
&
$ < 34 $
&
0
&
$ < 28 $
&
$ < 35 $
\\
\hline
$|\mu_{\nu_{e}}|$
&
$1$
&
$ < 36 $
&
$ < 52 $
&
$2$
&
$ < 45 $
&
$ < 59 $
\\
$|\mu_{\nu_{\mu}}|$
&
$16$
&
$ < 31 $
&
$ < 39 $
&
$22$
&
$ < 36 $
&
$ < 43 $
\end{tabular}
\caption{ \label{tab:all}
Best fits and limits at 90\% CL and 99\% CL for
the neutrino charge radii (in units of $10^{-32} \, \text{cm}^2$),
for the neutrino charges (in units of $10^{-8} \, e$),
and for the neutrino magnetic moments (in units of $10^{-10} \, \mu_{\text{B}}$).
}
\end{center}
\end{table*}

\begin{figure*}[!t]
\centering
\setlength{\tabcolsep}{0pt}
\begin{tabular}{cc}
\subfigure[]{\label{fig:chr7-rem-rmt}
\includegraphics*[width=0.3\linewidth]{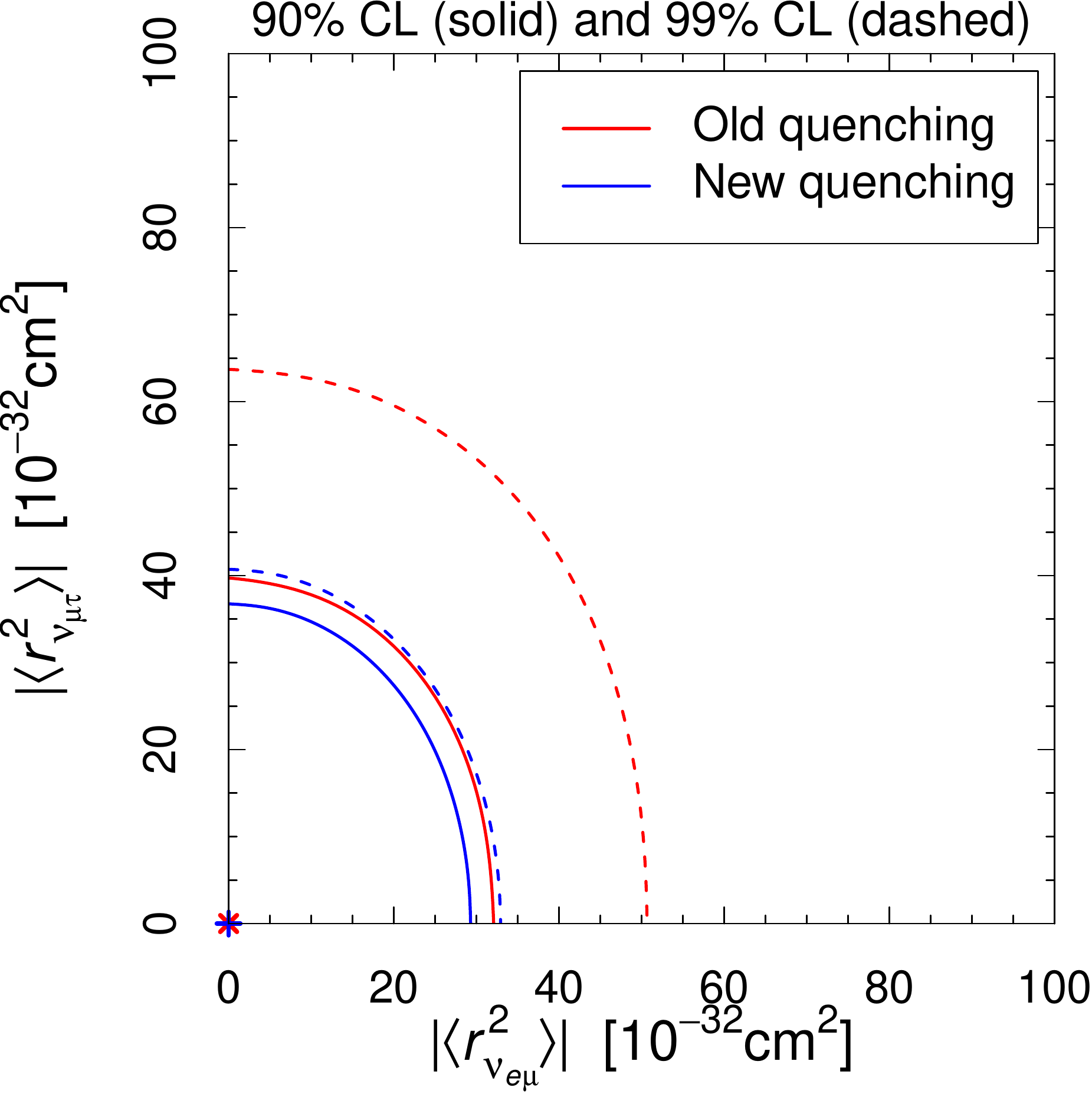}
}
&
\subfigure[]{\label{fig:chr7-ret-rmt}
\includegraphics*[width=0.3\linewidth]{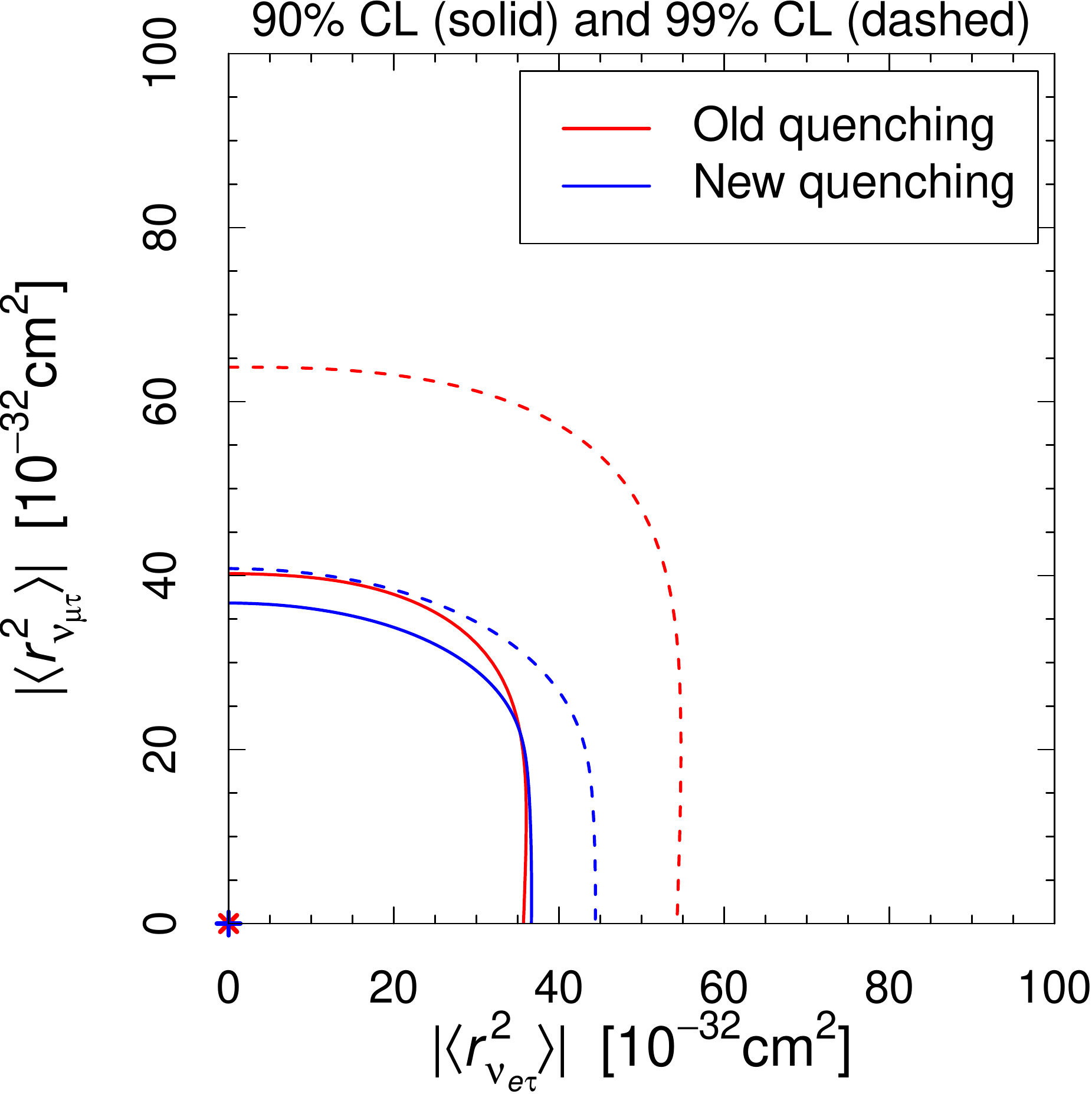}
}
\\
\subfigure[]{\label{fig:chr7-rem-ret}
\includegraphics*[width=0.3\linewidth]{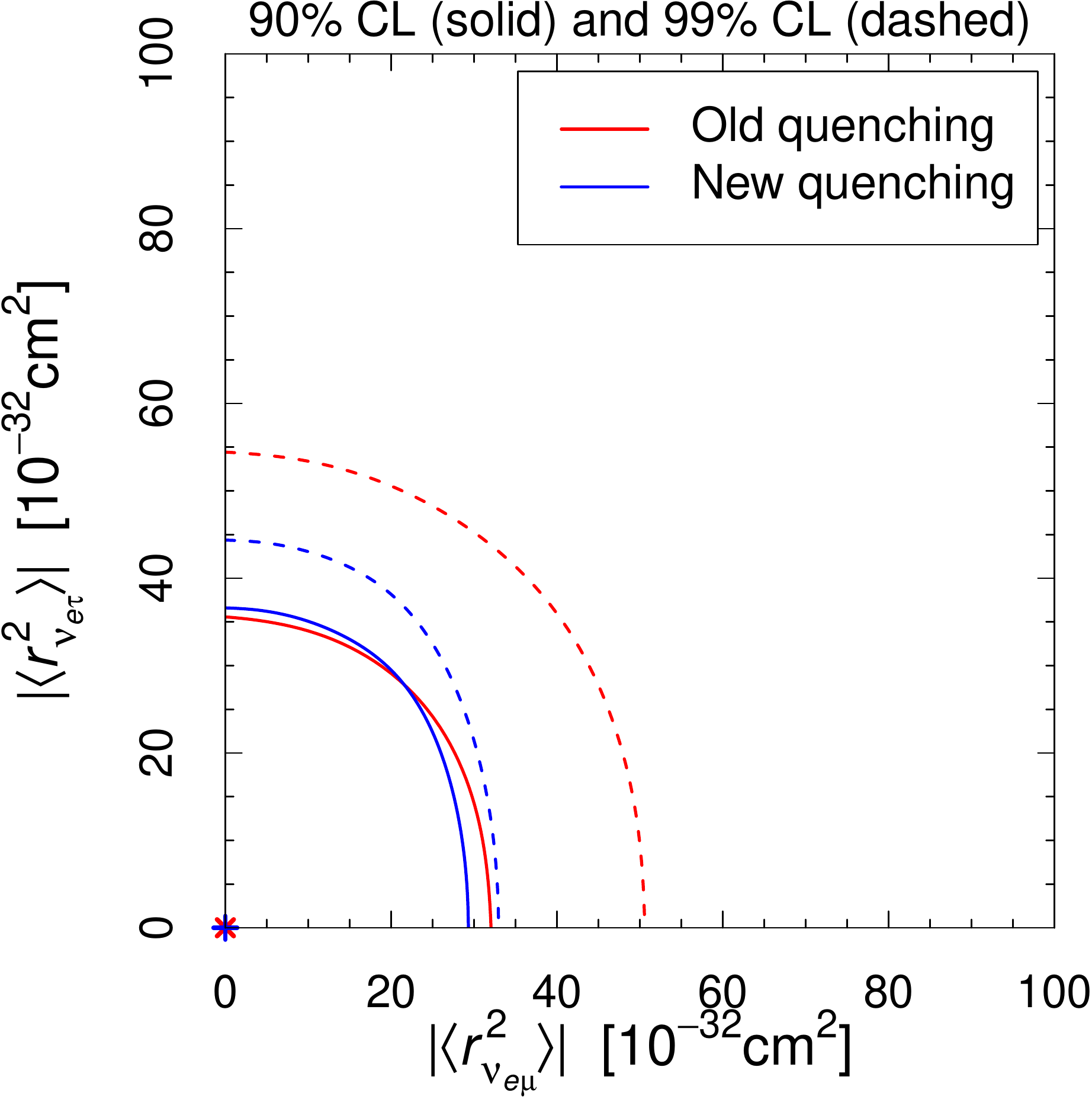}
}
&
\subfigure[]{\label{fig:chr7-ree-rmm}
\includegraphics*[width=0.3\linewidth]{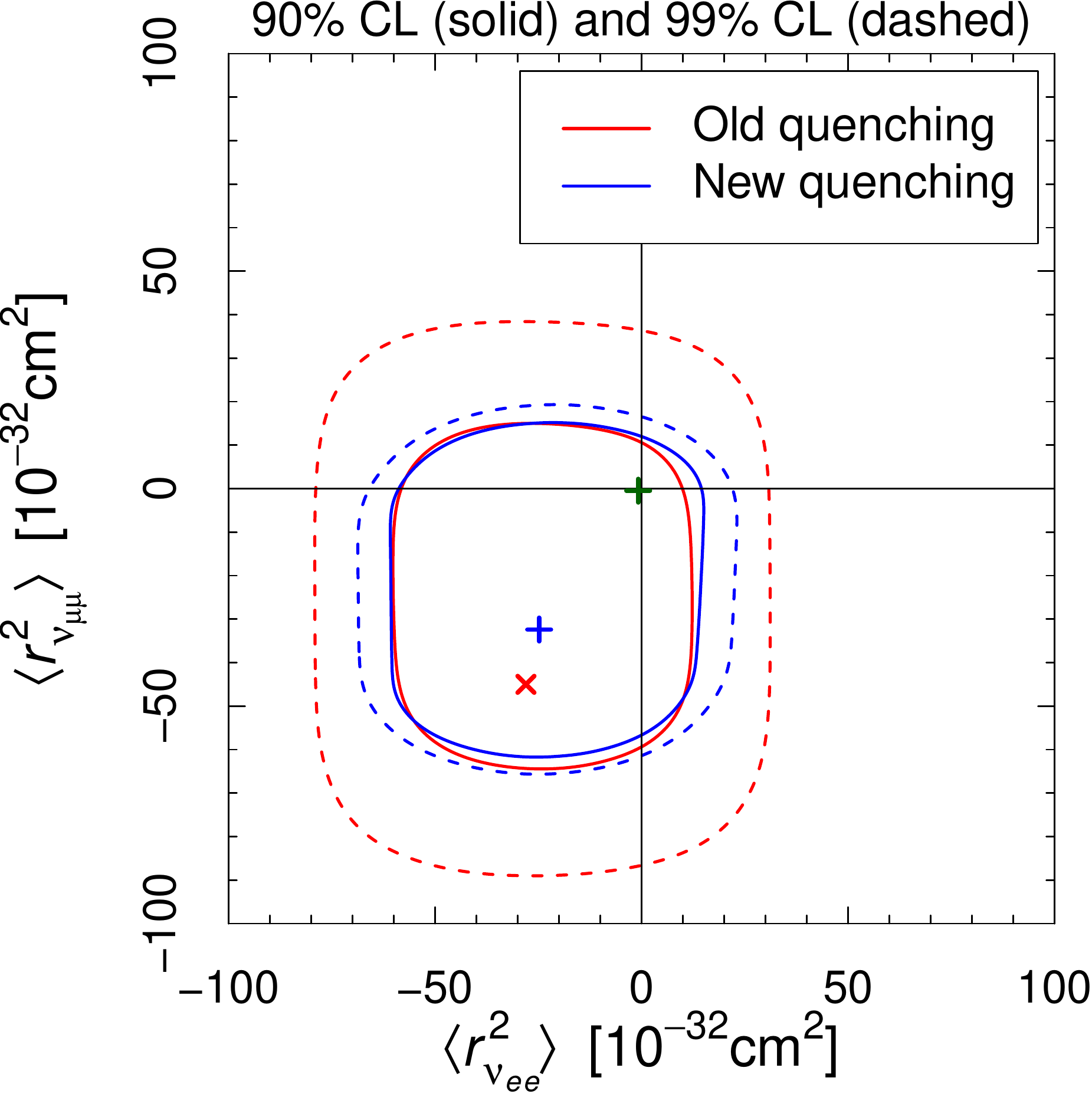}
}
\end{tabular}
\caption{ \label{fig:chr7}
Contours of the
90\% CL (solid blue curves) and 99\% CL (dashed blue curves)
allowed regions in different planes of the
neutrino charge radii parameter space
obtained with the new quenching factor~\protect\cite{Collar:2019ihs}
and free $R_{n}$.
For comparison,
also the corresponding contours obtained
with the old quenching factor~\protect\cite{Akimov:2017ade}
are shown in red.
The respective best-fit points are shown in blue and red.
The green point near the origin in panel \subref{fig:chr7-ree-rmm} indicates the Standard Model values
in Eqs.~(\ref{reSM}) and (\ref{rmSM}).
}
\end{figure*}

The results of our fits for fixed and free $R_{n}$ are given in Table~\ref{tab:all}.
One can see that the bounds are similar in the two cases.
Therefore, in Figure~\ref{fig:chr7}
we show only the allowed regions in different planes of the
neutrino charge radii parameter space obtained with free $R_{n}$
and compared with the corresponding allowed regions
obtained
with the old quenching factor~\protect\cite{Akimov:2017ade}.
One can see that there is
only a slight improvement of the 90\% CL allowed regions,
but the 99\% CL allowed regions are strongly reduced with the new quenching factor
and their contours lie close to those of the 90\% CL allowed regions.
Therefore,
the implementation of the new quenching factor
allows us to strengthen the statistical reliability of the bounds on the
neutrino charge radii at high confidence level values.

Our bounds on the neutrino charge radii are different with respect to
those obtained recently in Refs.~\cite{Papoulias:2019txv,Khan:2019mju}
with the new quenching factor
(taking into account that the charge radii in both papers are defined as half of ours).
Besides the reasons mentioned at the end of Section~\ref{sec:neutron},
as emphasized in the introduction and above,
our analysis is more powerful for flavor-dependent neutrino properties
because we take into account the arrival time information of the COHERENT data~\cite{Akimov:2018vzs},
that is not considered in Refs.~\cite{Papoulias:2019txv,Khan:2019mju}.
Moreover,
unfortunately the authors of Refs.~\cite{Papoulias:2019txv,Khan:2019mju}
adopted the incorrect treatment of the antineutrino charge radii
of Ref.~\cite{Cadeddu:2018dux}.

\begin{figure*}[!t]
\centering
\setlength{\tabcolsep}{0pt}
\begin{tabular}{cc}
\subfigure[]{\label{fig:chr2}
\includegraphics*[width=0.3\linewidth]{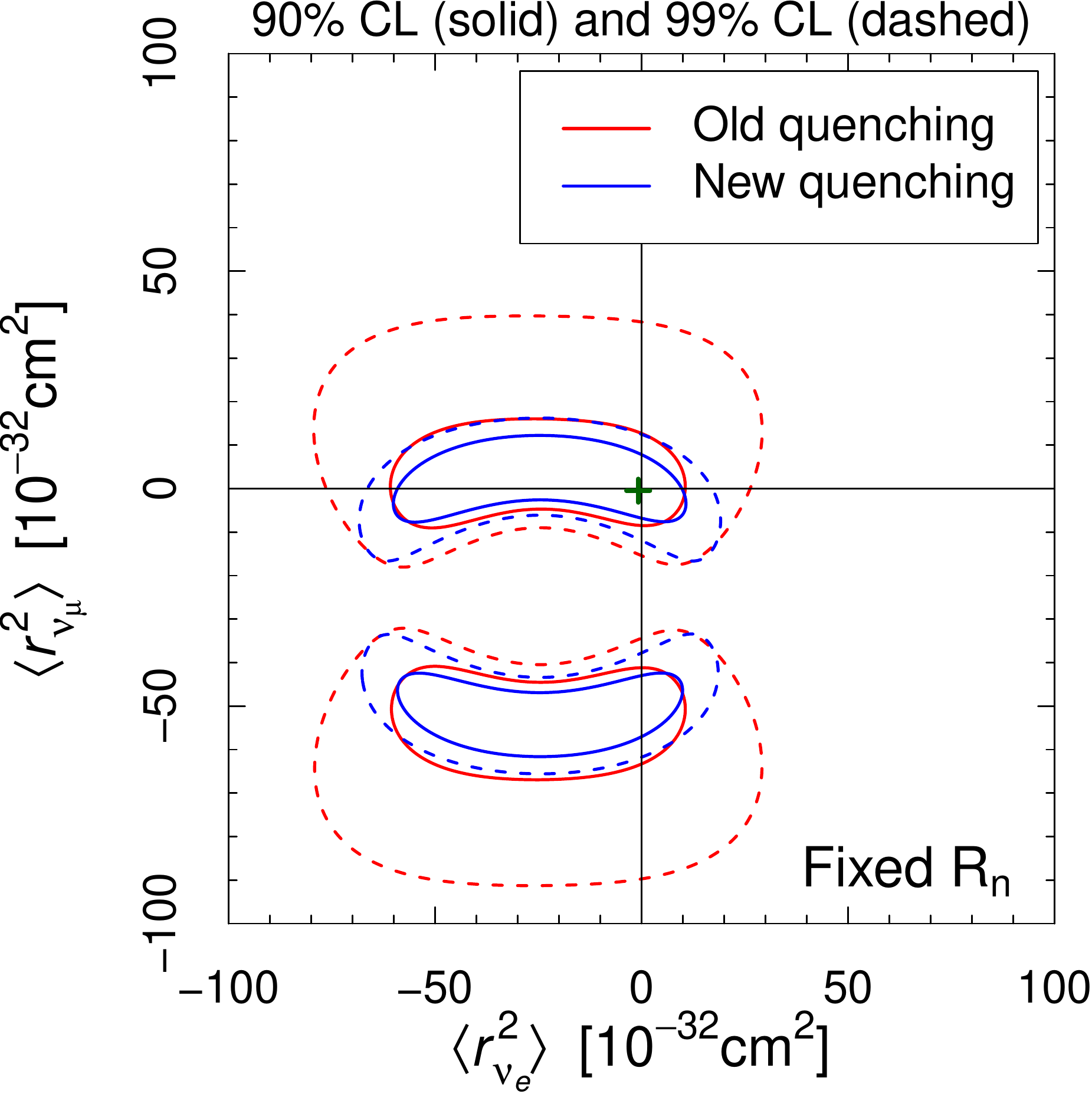}
}
&
\subfigure[]{\label{fig:chr4}
\includegraphics*[width=0.3\linewidth]{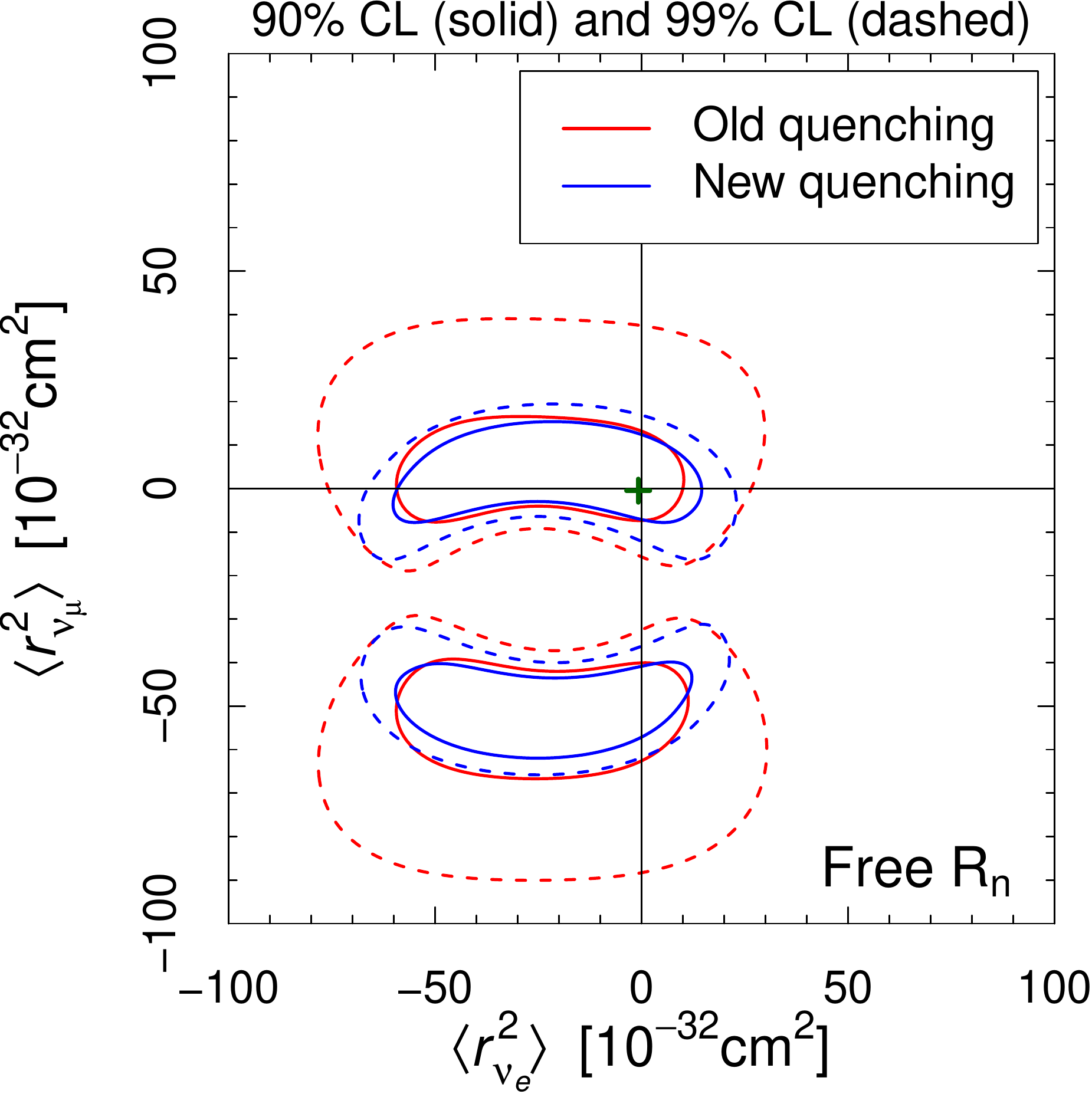}
}
\end{tabular}
\caption{ \label{fig:chr24}
Contours of the
90\% CL (solid blue curves) and 99\% CL (dashed blue curves)
allowed regions in the
($\langle{r}_{\nu_{e}}^2\rangle,\langle{r}_{\nu_{\mu}}^2\rangle$)
plane
obtained with the new quenching factor~\protect\cite{Collar:2019ihs}
and
\subref{fig:chr2} fixed, or
\subref{fig:chr4} free $R_{n}$,
assuming that the transition charge radii are negligible.
For comparison,
also the corresponding contours obtained
with the old quenching factor~\protect\cite{Akimov:2017ade}
are shown in red.
The green points near the origin indicate the Standard Model values
in Eqs.~(\ref{reSM}) and (\ref{rmSM}).
}
\end{figure*}

We can compare our bounds on the neutrino charge radii
with those obtained in Ref.~\cite{Kosmas:2017tsq}
using the old quenching factor~\protect\cite{Akimov:2017ade}
and considering only the total number of COHERENT events,
but with the correct treatment of
the sign of the contributions of the antineutrino charge radii.
The authors of Ref.~\cite{Kosmas:2017tsq}
considered only the diagonal charge radii
$\langle{r}_{\nu_{e}}^2\rangle \equiv \langle{r}_{\nu_{ee}}^2\rangle$
and
$\langle{r}_{\nu_{\mu}}^2\rangle \equiv \langle{r}_{\nu_{\mu\mu}}^2\rangle$.
This is an acceptable assumption
taking into account that in the Standard Model there are only diagonal charge radii,
as explained above.
Hence,
we present in Figure~\ref{fig:chr24}
the results of our analyses under this assumption,
considering the old and new quenchings
and
fixed and free $R_{n}$.
One can see that the allowed regions
are significantly reduced using the new quenching,
especially the one at 99\% CL.
Therefore, as in the general analysis with all the neutrino charge radii,
the implementation of the new quenching factor
leads to a strengthening of the statistical reliability of the bounds.
From the two panels in Figure~\ref{fig:chr24}
one can also see that the allowed regions of the diagonal charge radii
are rather similar for fixed and free $R_{n}$.
The allowed regions in Figure~\ref{fig:chr24}
are more stringent and have a different shape of that in
Figure~6 of Ref.~\cite{Kosmas:2017tsq},
because the consideration in Ref.~\cite{Kosmas:2017tsq}
of the total number of COHERENT events only
do not give any information on the difference of the properties of different neutrino flavors.
Therefore,
the allowed region in Figure~6 of Ref.~\cite{Kosmas:2017tsq}
is symmetric under the exchange
$\langle{r}_{\nu_{e}}^2\rangle \leftrightarrows \langle{r}_{\nu_{\mu}}^2\rangle$
and has a circular shape.
Our analysis, instead,
distinguishes the charge radii of $\nu_{e}$ and $\nu_{\mu}$,
leading to two allowed regions which
cover a relatively wide range of $\langle{r}_{\nu_{e}}^2\rangle$
and two narrower ranges of $\langle{r}_{\nu_{\mu}}^2\rangle$.
A common feature of the two analyses is the exclusion of an intermediate region around
$
\langle{r}_{\nu_{e}}^2\rangle
\simeq
\langle{r}_{\nu_{\mu}}^2\rangle
\simeq
- 25 \times 10^{-32} \, \text{cm}^2
$.
This feature can be understood by noting that the
contribution of the diagonal charge radii
in the cross section (\ref{cs-chr})
approximately cancel the weak neutral current contributions for
\begin{equation}
\langle{r}_{\nu_{\ell}}^2\rangle
\simeq
- \dfrac{ 3 \, N }{ 4 \, Z \, m_{W}^2 \sin^2\!\vartheta_{W} }
\simeq
- 26 \times 10^{-32} \, \text{cm}^2
,
\label{cancel}
\end{equation}
for CsI.
In this estimate we neglected $g_{V}^{p} \simeq 0.023$
and approximated
$(N/Z)_{^{133}\text{Cs}} \simeq (N/Z)_{^{127}\text{I}} \simeq 1.4$.
Around the values in Eq.~(\ref{cancel}) of the diagonal charge radii
the cross section is strongly suppressed and cannot fit the COHERENT data.
This suppression explains the hole in the middle of the allowed region
in Figure~6 of Ref.~\cite{Kosmas:2017tsq}
and
the corresponding excluded area in the two panels in Figure~\ref{fig:chr24}
obtained with our analysis.
Obviously,
this excluded area does not appear in Figure~\ref{fig:chr7},
because in the general analysis
the transition charge radii can compensate the suppression of the
part of the cross section which depends on the diagonal charge radii.

\section{Neutrino electric charges}
\label{sec:charges}

Coherent neutrino-nucleus elastic scattering is obviously sensitive
not only to the neutrino charge radii,
but also to the neutrino charges,
if neutrinos are not exactly neutral.
This possibility can occur in theories beyond the Standard Model
and has been considered in many experimental and theoretical studies
(see the review in Ref.~\cite{Giunti:2014ixa}).
Here we present for the first time the bounds on the neutrino charges
obtained from the analysis of the COHERENT data.
The analysis is similar to that concerning the neutrino charge radii
presented in Section~\ref{sec:radii},
with the replacement of $\tilde{Q}_{\ell\ell'}$
in Eq.~(\ref{Qchr})
with~\cite{Kouzakov:2017hbc}
\begin{equation}
\tilde{Q}_{\ell\ell'}
=
\dfrac{ 4 \, m_{W}^2 \sin^2\!\vartheta_{W} }{ q^2 }
\, q_{\nu_{\ell\ell'}}
=
\dfrac{ 2 \sqrt{2} \pi \alpha }{ G_{\text{F}} q^2 }
\, q_{\nu_{\ell\ell'}}
,
\label{Qech}
\end{equation}
where $ q^2 = - 2 M T $ is the squared four-momentum transfer.
As in the case of the charge radii,
although the charges of neutrinos and antineutrinos are opposite,
they contribute with the same sign to the shift of
$\sin^2\!\vartheta_{W}$,
because also the weak neutral current couplings change sign
from neutrinos to antineutrinos.

\begin{figure*}[!t]
\centering
\setlength{\tabcolsep}{0pt}
\begin{tabular}{cc}
\subfigure[]{\label{fig:ech7-qem-qmt}
\includegraphics*[width=0.3\linewidth]{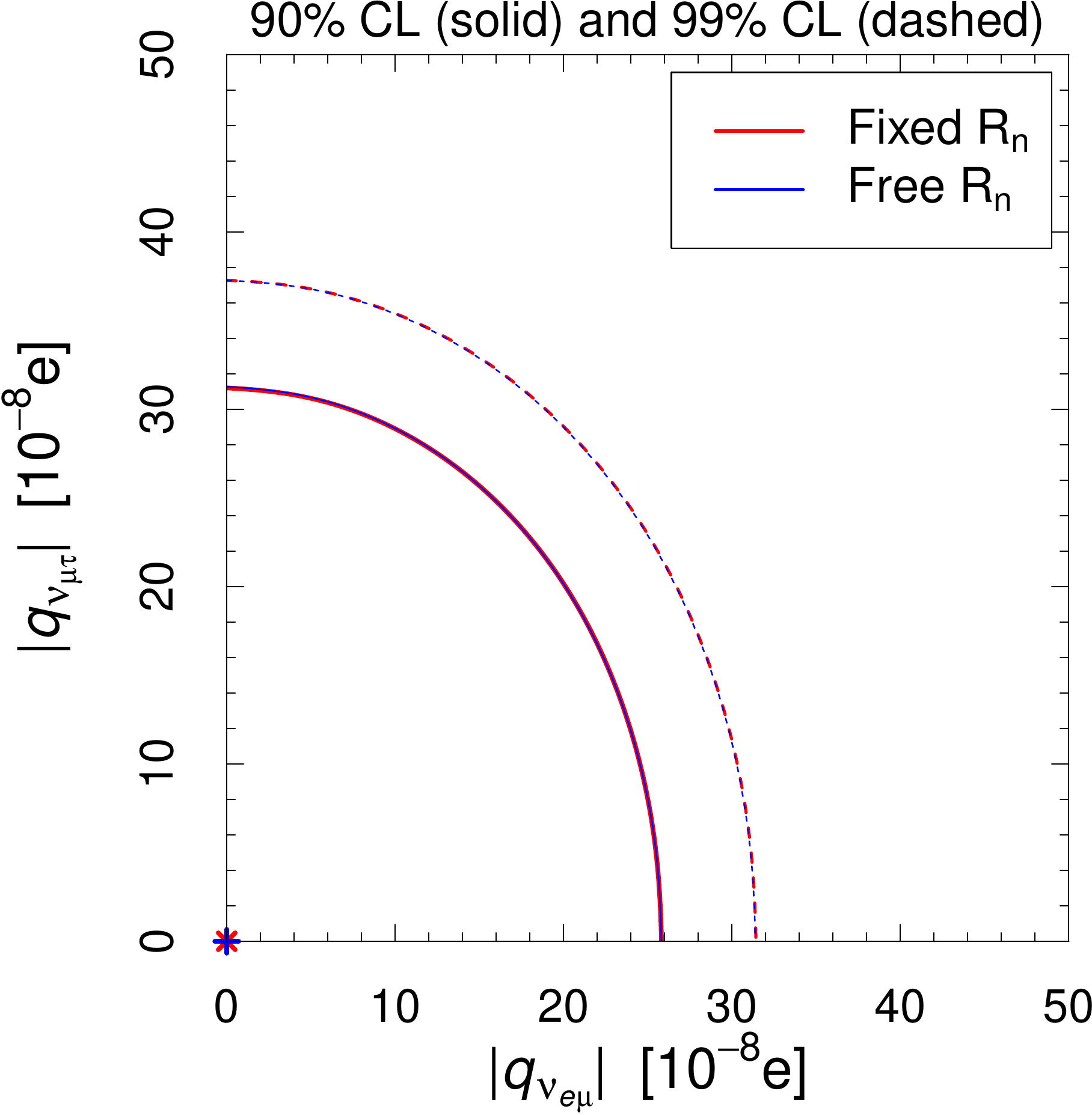}
}
&
\subfigure[]{\label{fig:ech7-qet-qmt}
\includegraphics*[width=0.3\linewidth]{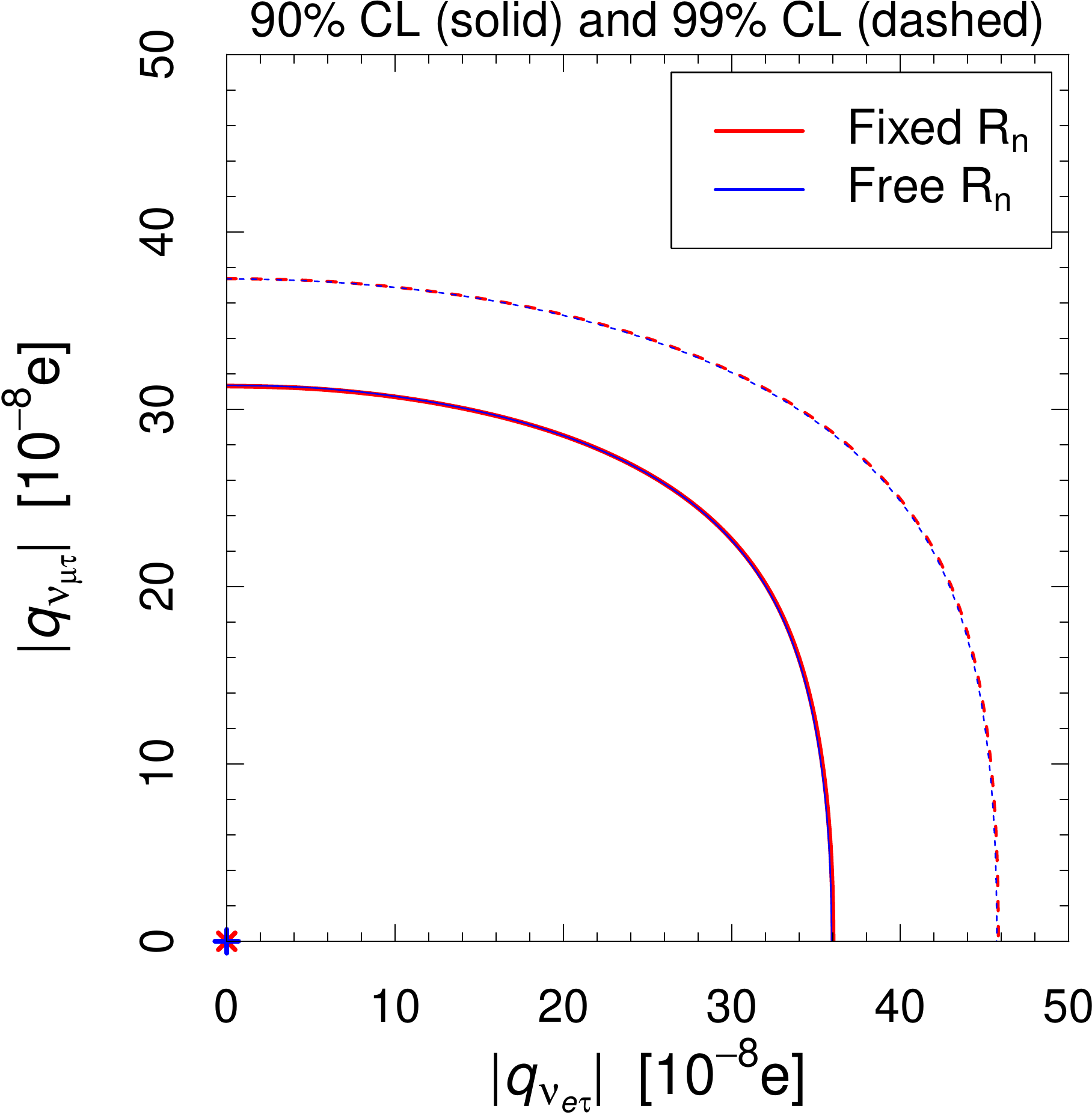}
}
\\
\subfigure[]{\label{fig:ech7-qem-qet}
\includegraphics*[width=0.3\linewidth]{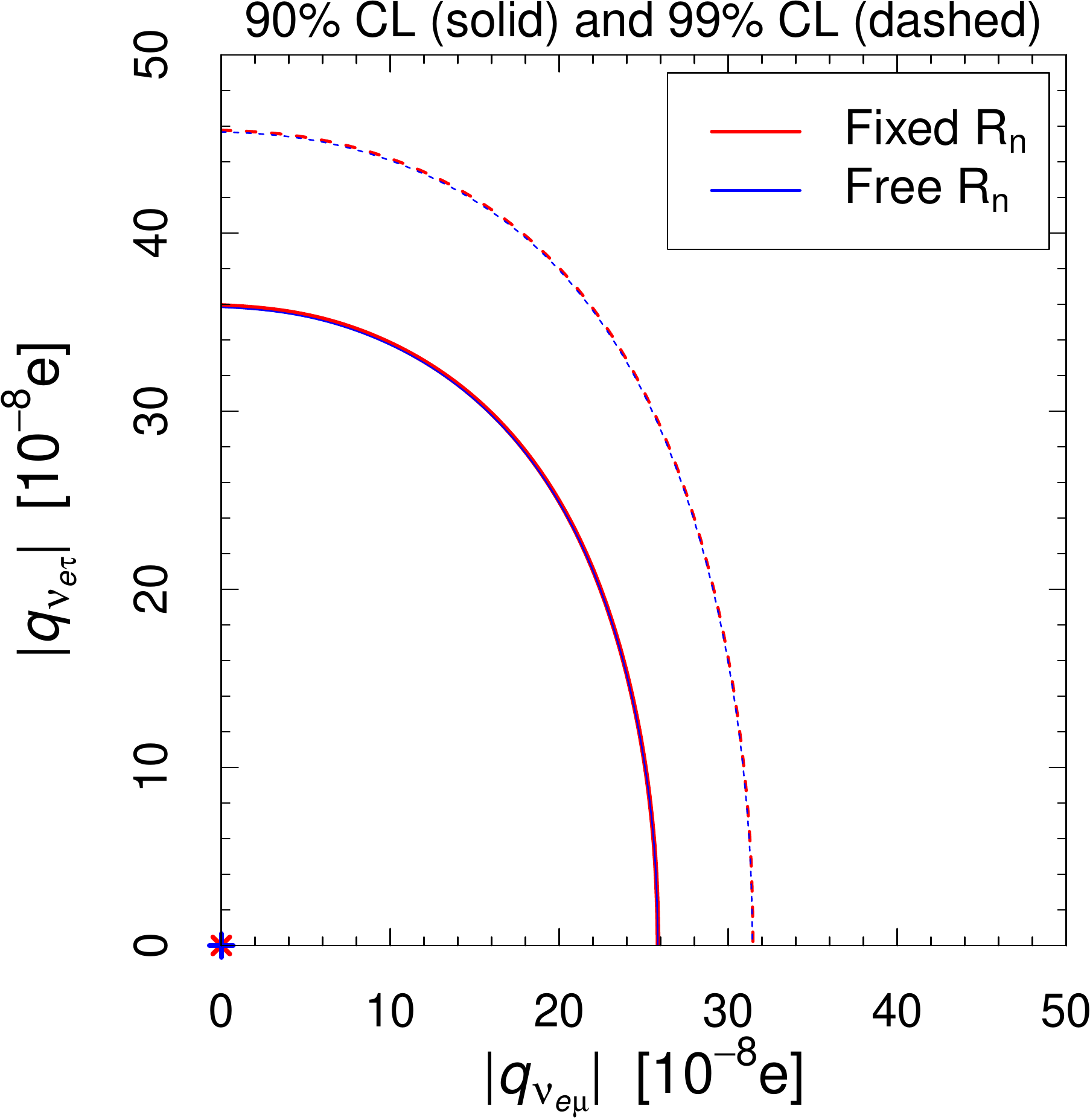}
}
&
\subfigure[]{\label{fig:ech7-qee-qmm}
\includegraphics*[width=0.3\linewidth]{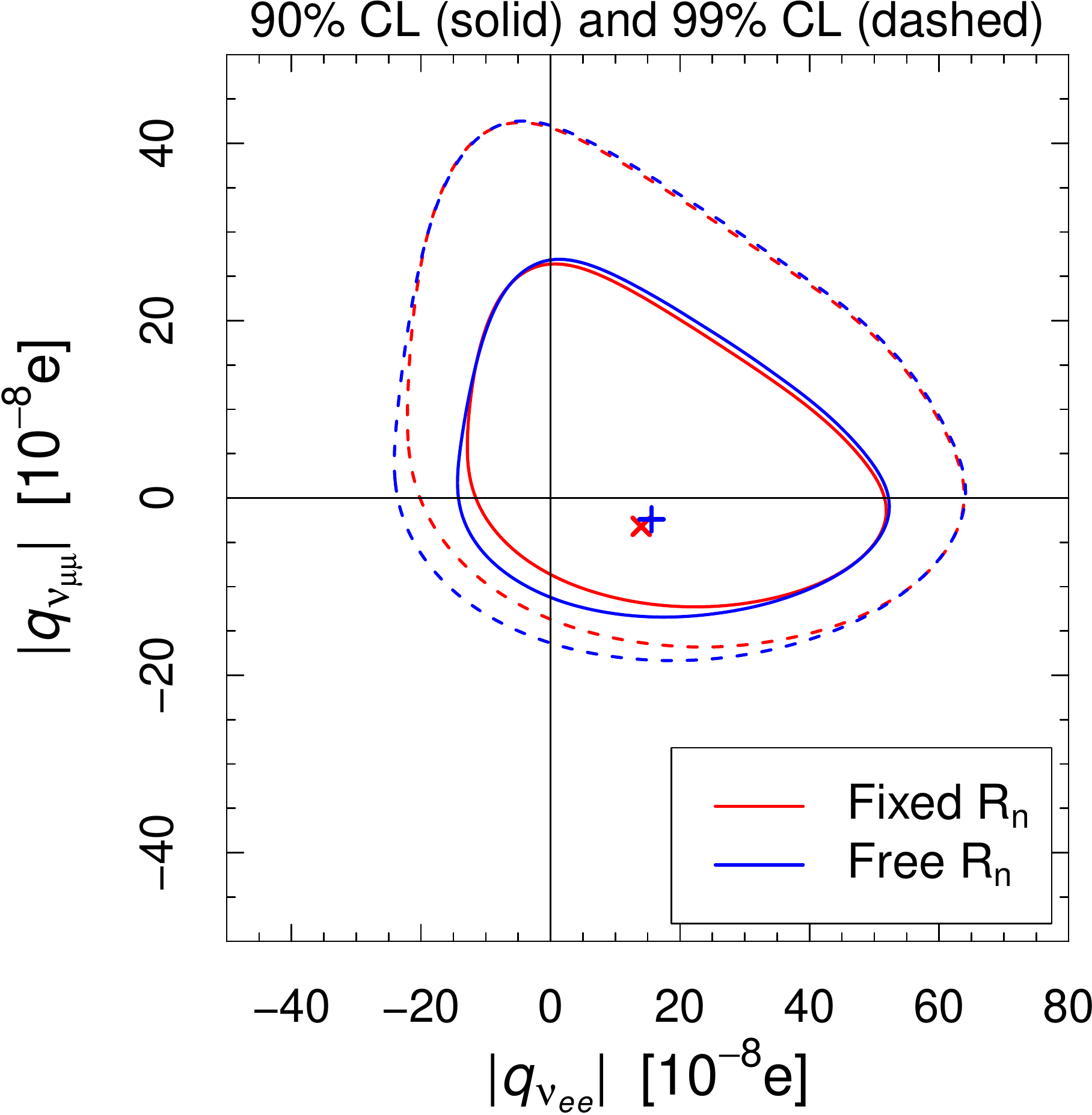}
}
\end{tabular}
\caption{ \label{fig:ech7}
Contours of the
90\% CL (solid curves) and 99\% CL (dashed curves)
allowed regions in different planes of the
neutrino electric charge parameter space
obtained with the new quenching factor~\protect\cite{Collar:2019ihs}
and with fixed (red) and free (blue) $R_{n}$.
The respective best-fit points are shown in blue and red.
}
\end{figure*}

The results of our fits for fixed and free $R_{n}$ are given in Table~\ref{tab:all}.
The allowed regions in different planes of the
neutrino electric charge parameter space are shown in Figure~\ref{fig:ech7}.
One can see that the bounds for all the neutrino charges are of the order of $10^{-7} \, e$.
Therefore the bounds on the charges involving the electron neutrino flavor
($q_{\nu_{ee}}$, $q_{\nu_{e\mu}}$, $q_{\nu_{e\tau}}$)
are not competitive with respect to those obtained in reactor neutrino experiments,
that are at the level of
$10^{-12} \, e$~\cite{Giunti:2014ixa,Chen:2014dsa}
for the effective charge
$\sqrt{ q_{\nu_{ee}}^2 + q_{\nu_{e\mu}}^2 + q_{\nu_{e\tau}}^2 }$
in neutrino-electron elastic scattering experiments.
On the other hand,
our bounds on $q_{\nu_{\mu\mu}}$ and $q_{\nu_{\mu\tau}}$
are the first ones obtained from laboratory data.

Let us comment
on an approximation in our cross sections
for electromagnetic neutrino-nucleus interactions.
Neglecting weak interactions and the nuclear form factor,
and considering only one generic neutrino charge $q_{\nu}$
(i.e. $ q_{\nu_{\ell\ell'}} = q_{\nu} \delta_{\ell\ell'} $),
from the cross section in Eq.~(\ref{cs-chr})
we obtain
\begin{equation}
\dfrac{d\sigma_{\nu\text{-}\mathcal{N}}^{(q_{\nu})}}{d T}
(E,T)
=
\dfrac{ 2 \pi \alpha^2 }{ M T^2 }
\left(
1 - \dfrac{M T}{2 E^2}
\right)
Z^2 q_{\nu}^2
.
\label{cs-charge}
\end{equation}
This is the cross section for the electric charge interaction of a fermion with charge $q_{\nu}$ with a point-like nucleus with $Z$ protons,
that can be obtained from the well-known Rosenbluth cross section
(see, for example Ref.~\cite{Alberico:2001sd})
for $T \ll E \ll M$,
neglecting the anomalous magnetic moment of the nucleus.
The omission of the effects due to the anomalous magnetic moment of the nucleus is an approximation of our calculations,
that is justified by the small contribution of the anomalous magnetic moments of
$^{133}\text{Cs}$ and $^{127}\text{I}$
with respect to their charges.
Indeed, the magnetic moments of the two nuclei are
$\mu(^{133}\text{Cs}) = 2.58 \, \mu_{\text{N}}$
and
$\mu(^{127}\text{I}) = 2.81 \, \mu_{\text{N}}$
(see Ref.~\cite{periodictable.com}),
and the Dirac magnetic moment of a point-like nucleus with electric charge $Z$ and atomic mass $A$ is given by $(Z/A) \, \mu_{\text{N}}$,
which gives
$\mu_{\text{Dirac}}(^{133}\text{Cs}) = 0.41 \, \mu_{\text{N}}$
and
$\mu_{\text{Dirac}}(^{127}\text{I}) = 0.42 \, \mu_{\text{N}}$.
Therefore,
the anomalous magnetic moments of
$^{133}\text{Cs}$ and $^{127}\text{I}$
are not enhanced with respect to the anomalous proton and neutron magnetic moments.
Taking also into account that the magnetic moment contribution to the cross section is suppressed at the low $q^2$ values that we are considering,
our approximation is well justified.

The fact that under the above approximations
we obtain the right equation (\ref{cs-charge})
implies the correctness of our
normalization of the charge radius that is twice
of that in Refs.~\cite{Papoulias:2019txv,Khan:2019mju}
and some other papers
(see the discussion in Ref.~\cite{Cadeddu:2018dux}).
One can see it by considering the sum of the values of $\tilde{Q}_{\ell\ell'}$ in Eqs.~(\ref{Qchr}) and (\ref{Qech}),
\begin{equation}
\tilde{Q}_{\ell\ell'}
=
\dfrac{ 2 \sqrt{2} \pi \alpha }{ G_{\text{F}} q^2 }
\left( q_{\nu_{\ell\ell'}} + \dfrac{q^2}{6} \, \langle{r}_{\nu_{\ell\ell'}}^2\rangle \right)
,
\label{Qtot}
\end{equation}
that corresponds to the standard expansion of the charge form factor
(see, for example Ref.~\cite{Giunti:2014ixa})
\begin{equation}
F^{\nu}_{Q}(q^{2})
=
F^{\nu}_{Q}(0)
+
q^{2}
\left.\frac{d F^{\nu}_{Q}(q^{2})}{d q^{2}}\right|_{q^{2}=0}
+
\ldots
=
q_{\nu}
+
\dfrac{ q^{2} }{6} \, \langle r_{\nu}^2 \rangle
+
\ldots
.
\label{ff}
\end{equation}
As we have seen above, these relations lead to the
correct cross section (\ref{cs-charge}) for the electric charge interaction of a fermion with charge $q_{\nu}$ with a point-like nucleus with $Z$ protons.
If instead the normalization of the charge radius is half of ours,
the expression of the cross section is multiplied by a factor of four
and the standard relation between $q_{\nu}$ and $\langle r_{\nu}^2 \rangle$
in Eqs.~(\ref{Qtot}) and (\ref{ff})
leads to a cross section for the electric charge interaction of a fermion with charge $q_{\nu}$ with a point-like nucleus with $Z$ protons
that is four times larger than the correct one in Eq.~(\ref{cs-charge}).

\section{Neutrino magnetic moments}
\label{sec:magnetic}

The COHERENT data on coherent neutrino-nucleus elastic scattering
have been also analyzed taking into account the effects of possible
neutrino magnetic moments~\cite{Kosmas:2017tsq,Papoulias:2019txv,Khan:2019mju}.
In this Section we present our bounds on the neutrino magnetic moments
taking into account the new quenching factor
and the arrival time information of the COHERENT data,
that was not considered in Refs.~\cite{Kosmas:2017tsq,Papoulias:2019txv,Khan:2019mju}.

For the analysis of the coherent data we use the least-squares function in Eq.~(\ref{chi2time}),
with the theoretical predictions $N_{ij}^{\text{th}}$
calculated by adding to the Standard Model weak cross section in Eq.~(\ref{cs-std})
the magnetic moment interaction cross section
\begin{equation}
\dfrac{d\sigma_{\nu_{\ell}\text{-}\mathcal{N}}^{\text{mag}}}{d T}
(E,T)
=
\dfrac{ \pi \alpha^2 }{ m_{e}^2 }
\left( \dfrac{1}{T} - \dfrac{1}{E} \right)
Z^2 F_{Z}^2(|\vet{q}|^2)
\left( \dfrac{\mu_{\nu_{\ell}}}{\mu_{\text{B}}} \right)^2
,
\label{cs-mag}
\end{equation}
where $m_{e}$ is the electron neutrino mass
and
$\mu_{\nu_{\ell}}$
is the effective magnetic moment of the flavor neutrino $\nu_{\ell}$
in elastic scattering
(see Ref.~\cite{Giunti:2014ixa}).

\begin{figure*}[!t]
\centering
\includegraphics*[width=0.5\linewidth]{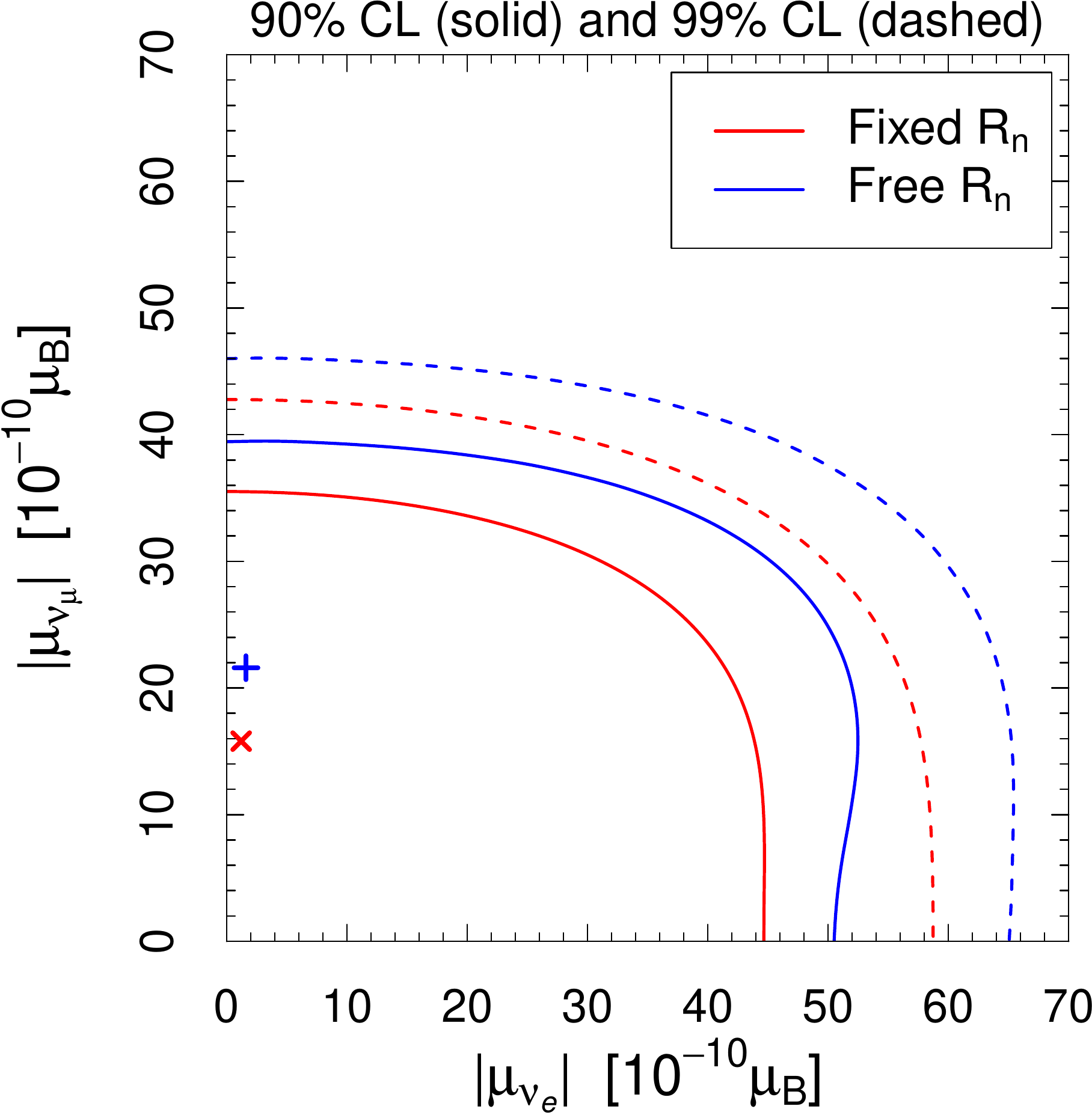}
\caption{ \label{fig:mag4}
Contours of the
90\% CL (solid curves) and 99\% CL (dashed curves)
allowed regions in the
($|\mu_{\nu_{e}}|,|\mu_{\nu_{\mu}}|$) plane
obtained with the new quenching factor~\protect\cite{Collar:2019ihs}
and with fixed (red) and free (blue) $R_{n}$.
The respective best-fit points are shown in blue and red.
}
\end{figure*}

The results of the fits for fixed and free $R_{n}$ are given in Table~\ref{tab:all}
and Figure~\ref{fig:mag4}.
One can see that the determination of the neutrino magnetic moments
is of the order of $10^{-9} \, \mu_{\text{B}}$,
with slightly more stringent constraints on $|\mu_{\nu_{\mu}}|$
with respect to $|\mu_{\nu_{e}}|$.
Unfortunately the sensitivity to $|\mu_{\nu_{e}}|$
is not competitive with that of reactor experiments,
that constrain $|\mu_{\nu_{e}}|$ at the level of $10^{-11} \, \mu_{\text{B}}$
\cite{Beda:2012zz,Giunti:2014ixa}.
On the other hand,
the best current laboratory limit on $|\mu_{\nu_{\mu}}|$
is $6.8 \times 10^{-10} \, \mu_{\text{B}}$ at 90\% CL
\cite{Auerbach:2001wg,Giunti:2014ixa},
that is only about 5 times smaller than our limit.

Our bounds on the neutrino magnetic moments are different from
those obtained recently in Refs.~\cite{Papoulias:2019txv,Khan:2019mju}
for the reasons mentioned at the end of Sections~\ref{sec:neutron} and \ref{sec:radii}.
Moreover our bounds are more stringent than those found in Ref.~\cite{Khan:2019mju}.
On the other hand, they are slightly less stringent than the bound
found in Ref.~\cite{Papoulias:2019txv}
without distinguishing between $|\mu_{\nu_{e}}|$ and $|\mu_{\nu_{\mu}}|$.

\section{Conclusions}
\label{sec:conclusions}

In this paper we updated the analyses of the
coherent neutrino-nucleus elastic scattering data
of the COHERENT experiment~\cite{Akimov:2017ade}
presented in Refs.~\cite{Cadeddu:2017etk,Cadeddu:2018izq,Cadeddu:2018dux}
taking into account the new quenching factor
published recently in Ref.~\cite{Collar:2019ihs}.
These updates lead to better determinations of the average rms radius of the neutron distributions
of $^{133}\text{Cs}$ and $^{127}\text{I}$,
of the low-energy weak mixing angle,
and of the neutrino charge radii.
We also presented interesting new constraints on the neutrino charges
and on the neutrino magnetic moments.

The new determination of the CsI neutron distribution radius
is significantly improved with respect to that in Ref.~\cite{Cadeddu:2017etk},
with smaller uncertainties and a best fit value that is in agreement
with nuclear model predictions.
We also improved the evidence of the suppression of coherence due to the nuclear structure
from $2.3\sigma$ of Ref.~\cite{Cadeddu:2017etk}
to
$3.7\sigma$.

The updated determination of the weak mixing angle using the APV measurement is much more precise with respect to that in Ref.~\cite{Cadeddu:2018izq}, and despite the changed value of the $^{133}\text{Cs}$ neutron distribution radius used as an input, thanks to the usage of a new and more precise determination of the vector transition polarizability~\cite{Toh:2019iro} the result is still in very good agreement with the Standard Model. The combined fit of the APV and COHERENT data allows moreover to obtain an even more precise determination of the $^{133}\text{Cs}$ neutron distribution radius and neutron skin. 

In the analysis of the COHERENT data considering the effects of the neutrino charge radii, we corrected the treatment of the sign of the contributions of the antineutrino charge radii
in Ref.~\cite{Cadeddu:2018dux}
(see the discussion in Section~\ref{sec:radii}).
We have shown that the new quenching factor leads to a significant improvement of the
constraints on the charge radii,
especially at high values of the confidence level.
This allows us to strengthen the statistical reliability of the bounds on the
neutrino charge radii.

Our constraints on the neutrino charges are the first ones obtained from
coherent neutrino-nucleus elastic scattering.
Unfortunately,
the bounds on the charges involving the electron neutrino flavor
($q_{\nu_{ee}}$, $q_{\nu_{e\mu}}$, $q_{\nu_{e\tau}}$)
are not competitive with respect to those obtained in reactor neutrino experiments,
being about five orders of magnitude larger.
On the other hand,
the bounds on
the diagonal charge $q_{\nu_{\mu\mu}}$ of $\nu_{\mu}$ and
the $\nu_{\mu}$-$\nu_{\tau}$ transition charge $q_{\nu_{\mu\tau}}$
are the first ones obtained from laboratory data.

Our constraints on the effective electron neutrino magnetic moment $|\mu_{\nu_{e}}|$ are not competitive with the current reactor limits,
that are about two orders of magnitude better,
but
our constraints on $|\mu_{\nu_{\mu}}|$
are only about 5 times larger than the best current laboratory limits.

We have also commented on the differences of our analysis and results
with respect to those presented recently in Refs.~\cite{Papoulias:2019txv,Khan:2019mju},
that used the new quenching factor in Ref.~\cite{Collar:2019ihs}.
The main sources of differences are the fit of only the total number of events in
Ref.~\cite{Papoulias:2019txv}
and
the fit of only the COHERENT energy spectrum with a constant quenching factor
in Ref.~\cite{Khan:2019mju}.
Instead, in our analysis we have used the energy-dependent quenching factor in Ref.~\cite{Collar:2019ihs}
and we have analyzed the time- and energy-dependent COHERENT data,
that allow a better discrimination between the properties of
$\nu_{e}$ and $\nu_{\mu}$.

\begin{acknowledgments}
C.G. would like to thank Arun Thalapillil for stimulating discussions on neutrino millicharges.
The work of Y.F.Li and Y.Y. Zhang is supported by the National Natural Science Foundation of China under Grant No. 11835013, by the Strategic Priority Research Program of the Chinese Academy of Sciences under Grant No. XDA10010100. Y.F. Li is also grateful for the support by the CAS Center for Excellence in Particle Physics (CCEPP).
\end{acknowledgments}

%

\end{document}